\documentclass[aps,superscriptaddress,notitlepage,reprint]{revtex4-2}
\usepackage{epsfig,color}
\usepackage{graphicx}
\usepackage{dcolumn}
\usepackage{bm}
\usepackage{amsmath,amsfonts,amssymb,mathrsfs}
\usepackage{pstricks}
\usepackage{amsxtra}
\usepackage{cases}
\usepackage{amsthm}
\usepackage{braket}
\usepackage{natbib}
\usepackage{physics}
\usepackage{hyperref,color}
\usepackage{lipsum}
\usepackage{comment}
\definecolor{darkblue}{rgb}{0.0,0.0,0.7}
\hypersetup{colorlinks,breaklinks,linkcolor=darkblue,urlcolor=darkblue,anchorcolor=darkblue,citecolor=darkblue}

\begin{document}

\title{Scalable Suppression of XY Crosstalk by Pulse-Level Control in \\Superconducting Quantum Processors}

\author{Hui-Hang Chen}
\affiliation{Department of Physics and Center for Theoretical Physics, National Taiwan University, Taipei 106319, Taiwan}

\author{Chiao-Hsuan Wang}
\email{chiaowang@phys.ntu.edu.tw}
\affiliation{Department of Physics and Center for Theoretical Physics, National Taiwan University, Taipei 106319, Taiwan}
\affiliation{Center for Quantum Science and Engineering, National Taiwan University, Taipei 106319, Taiwan}
\affiliation{Physics Division, National Center for Theoretical Sciences, Taipei, 106319, Taiwan}

\begin{abstract}
As superconducting quantum processors continue to scale, high-performance quantum control becomes increasingly critical. In densely integrated architectures, unwanted interactions between nearby qubits give rise to crosstalk errors that limit operational performance. In particular, direct exchange-type (XY) interactions are typically minimized by designing large frequency detunings between neighboring qubits at the hardware level. However, frequency crowding in large-scale systems ultimately restricts the achievable frequency separation. While such XY coupling facilitates entangling gate operations, its residual presence poses a key challenge during single-qubit controls. Here, we propose a scalable pulse-level control framework, incorporating frequency modulation (FM) and dynamical decoupling (DD), to suppress XY crosstalk errors. This framework operates independently of coupling strengths, reducing calibration overhead and naturally supporting multi-qubit connectivity. Numerical simulations show orders-of-magnitude reductions in infidelity for both idle and single-qubit gates in a two-qubit system. We further validate scalability in a five-qubit layout, where crosstalk between a central qubit and four neighbors is simultaneously suppressed. Our crosstalk suppression framework provides a practical route toward high-fidelity operation in dense superconducting architectures.
\end{abstract}

\maketitle

\section{Introduction}
Enabled by advances in microwave-based control and flexible circuit architecture, superconducting qubit systems have become one of the leading platforms for quantum computation~\cite{Wendin2017,Krantz2019,Kjaergaard2020,Blais2021}.
As superconducting quantum processors have reached the thousand-qubit regime~\cite{Castelvecchi2023}, achieving high-fidelity operations within large-scale systems has become increasingly critical. Recent demonstrations of quantum error correction~\cite{Acharya2025} further indicate that practical fault-tolerant operation may be within reach~\cite{Shor1996,Fowler2012}. These developments highlight the importance of enhancing the performance of quantum hardware to reliably operate below the fault-tolerance thresholds for next-generation superconducting architectures~\cite{Knill2005}.
\vspace{1em}

Unintended effects associated with neighboring qubits have become an increasingly prominent source of error in large-scale quantum processors~\cite{Sheldon2016,McKay2019,Zhao2022,Ketterer2023}. Such crosstalk errors not only degrade gate and measurement performance, but also generate nonlocal error patterns that are particularly detrimental to quantum error correction~\cite{vonLupke2020,Sarovar2020}. One major mechanism, commonly referred to as classical crosstalk~\cite{Zhao2022}, arises from microwave or flux-control pulses unintentionally driving neighboring qubits during active operations~\cite{Sheldon2016,Abrams2019}. These control-induced errors can be reduced by improving hardware isolation~\cite{Heinsoo2018,Kosen2024} and further mitigated through pulse-shaping or compensation techniques~\cite{Wang2022,Aguila2025}. Another significant mechanism, often termed quantum crosstalk~\cite{Zhao2022}, stems from static interactions between nearby qubits, which perturb qubit evolution during both active operations and idle periods.

We focus on the quantum crosstalk arising from XY interactions between neighboring qubits. Such interactions, typically originating from capacitive or inductive coupling, are well described by transverse exchange (XY-type) interactions~\cite{Krantz2019,Blais2021}, enabling iSWAP-type two-qubit entangling gates~\cite{Majer2007,Han2020}.
When neighboring qubits are strongly detuned, direct excitation exchange is suppressed, leaving effective ZZ-type interactions and frequency shifts as the dominant residual effects~\cite{DiCarlo2009,Magesan2020,Fors2024}. Consequently, mitigation strategies have mainly targeted ZZ crosstalk~\cite{Mundada2019,Zhao2020,Ku2020,Li2020,Han2020,Zhao2021,Sung2021,Stehlik2021,Kandala2021,Tripathi2022,Liang2024,Niu2024}. However, as processors scale and spectral crowding becomes increasingly severe, maintaining large detuning across all neighboring qubits is no longer feasible, and XY coupling emerges as a significant source of coherent crosstalk error in dense superconducting architectures.

In this work, we develop pulse-level control strategies to suppress residual XY coupling during single-qubit operations without additional hardware overhead. We first show that XY-induced errors can be passively reduced by choosing gate durations that average out the unwanted exchange interaction. Building on this idea, we propose active approaches based on engineering the instantaneous detuning between qubits, including continuous frequency modulation (FM) and discrete dynamical-decoupling (DD) sequences, both of which further suppress XY crosstalk while remaining compatible with single-qubit gates. Finally, we validate the scalability of these methods in a five-qubit system, highlighting their applicability to larger superconducting processors.

\section{Crosstalk Dynamics}
\begin{figure}[htbp]
\begin{center}
\includegraphics[width=0.7\linewidth]{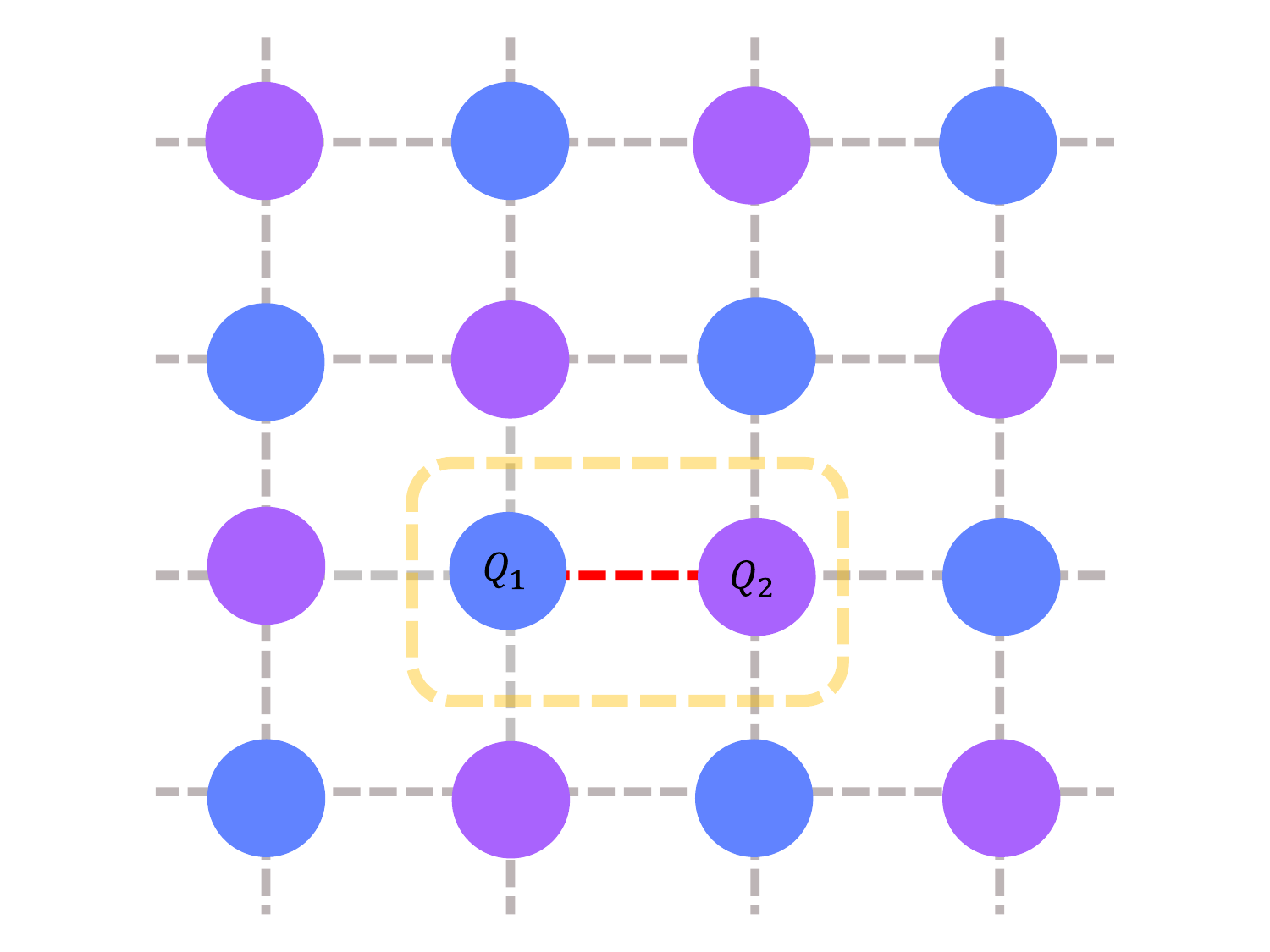}
    \caption{Schematic of a nearest-neighbor qubit array with XY interactions, indicated by gray dashed lines. The pair $Q_1$ and $Q_2$ within the dashed orange outline defines a minimal two-qubit subsystem used to analyze XY-induced crosstalk under single-qubit control.
The XY coupling between $Q_1$ and $Q_2$ is highlighted by red dashed lines.}
    \label{fig:Schematics}
\end{center}
\end{figure}

We consider two superconducting qubits coupled through an exchange-type interaction, within the 2D lattice layout, as shown in Fig.~\ref{fig:Schematics}. The coupling is represented by the XY Hamiltonian
\begin{align}
\hat{H}_{\rm XY} = J \left( \hat{\sigma}_1^{+} \hat{\sigma}_2^{-} + \hat{\sigma}_1^{-} \hat{\sigma}_2^{+} \right) =  \frac{J}{2} \left( \hat{\sigma}_1^{x} \hat{\sigma}_2^{x} + \hat{\sigma}_1^{y} \hat{\sigma}_2^{y} \right),
\end{align}
where $J$ is the coupling strength and $\hat{\sigma}_{j}^{\pm} = (\hat{\sigma}_{j}^{x} \pm i\hat{\sigma}_{j}^{y})/2$ are the raising and lowering operators of qubit $j$, with $\hat{\sigma}_{j}^{x,y,z}$ denoting the Pauli matrices acting on its computational subspace.  We take $\hbar=1$ throughout. This term describes coherent excitation exchange between adjacent qubits, which enables two-qubit entangling gates but also leads to unwanted interactions during single-qubit operations. In particular, when controlling one qubit while the neighboring qubit remains idle, residual XY coupling can induce population leakage and phase errors, reducing gate fidelity. Suppressing these errors while preserving the intended interactions is crucial for high-fidelity quantum computation.

Our goal is to develop an enhanced control scheme capable of suppressing the errors induced by residual XY coupling during single-qubit and idle operations. During gate operations, the system evolution under crosstalk dynamics (CD) is governed by the total Hamiltonian in the laboratory frame, given by
\begin{align}
    \hat{H}^{\rm CD}(t) &= \hat{H}_0 +  \hat{H}_{\rm XY} + \hat{H}_{\rm drive}(t).
    \label{eqn:HCD}
\end{align}
where $\hat{H}_0=-\sum_{i=1}^{2}\frac{\omega_{i}}{2}\hat{\sigma}_{i}^z$ is the bare qubit Hamiltonian and $\omega_{i}$ is the frequency of the qubit $i$, and $\hat{H}_{\rm {drive}}(t)$ represents the driving field applied to the qubits to achieve the target operations.

To understand how the XY interaction influences the computational basis, we move to the rotating frame defined by the qubit frequencies, which corresponds to the operation frame used for qubit state control and measurement~\cite{Krantz2019}. The transformation to this frame is defined by the unitary operator
\begin{align}
    \hat{U}_{H_0}(t) = \exp\left[i \left( \frac{\omega_{1}}{2}\hat{\sigma}_{1}^z+\frac{\omega_{2}}{2}\hat{\sigma}_{2}^z\right) t \right],
    \label{eqn:rotQubit}
\end{align}
with the corresponding Hamiltonian in the rotating frame given by
\begin{align} 
    \tilde{H}^{\rm CD}(t) &= \hat{U}^{\dagger}_{H_0}(t)\hat{H}^{\rm CD}(t)\hat{U}_{H_0}(t) + i\Dot{\hat{U}}_{H_0}^{\dagger}(t)\hat{U}_{H_0}(t) \notag\\
    &=  \tilde{H}_{\rm drive}(t) + \tilde{H}_{\rm XY}(t).
    \label{eqn:HVCD}
\end{align}
The gate unitary in the presence of XY crosstalk is
\begin{align}
     \hat{U}^{\rm CD}_{\rm gate}(T,0) =  \mathcal{T}\exp \left\{ -i\int_{0}^{T} \left[\tilde{H}_{\rm drive}(t)+ \tilde{H}_{\rm XY}(t) \right] \mathrm{d}t \right\},
\label{eqn:UCDgate}
\end{align}
where $\mathcal{T}$ denotes the time ordering.  The unwanted XY term interferes with the intended control evolution, giving rise to crosstalk-induced gate errors. 

Using the Magnus expansion~\cite{Blanes2009}, the evolution operator can be expressed as
\begin{align}
\hat{U}^{\rm CD}_{\rm gate}(T,0) &= e^{-iT \left[\bar{H}^{\mathrm{CD}(1)}_{\mathrm{gate}} + \bar{H}^{\mathrm{CD}(2)}_{\mathrm{gate}}+\cdots \right] },
\label{eqn:UCDmagnus}
\end{align}
where $\bar{H}^{\mathrm{CD}(n)}_{\mathrm{gate}}$ denotes the $n$-th order term in the Magnus expansion, corresponding to the $n$-th order contributions to the effective time-averaged Hamiltonian.

Considering only the leading-order contribution, the effective first-order Hamiltonian associated with the XY interaction is
\begin{align}
    \bar{H}^{\mathrm{CD}(1)}_{\rm XY}&= \frac{1}{T} \int_{0}^{T}  \tilde{H}_{XY} (t)\mathrm{d}t.
\end{align}
The XY interaction in the rotating frame takes the form
\begin{align}
     \tilde{H}_{\rm XY}(t) = J( e^{i \Delta t }\hat{\sigma}_{1}^{+}\hat{\sigma}_{2}^{-} +  e^{-i \Delta t }\hat{\sigma}_{1}^{-}\hat{\sigma}_{2}^{+}).
\end{align}
The exchange term thus acquires an oscillatory phase factor dependent upon the detuning. This oscillatory behavior naturally suggests a timing-based mechanism for error suppression: by tailoring the phase accumulated through detuning, the effective exchange interaction can be averaged out over the gate duration.

In the simplest case of static qubit frequencies, XY error suppression can be achieved by choosing specific gate durations.
Averaging the interaction over the gate duration yields
\begin{align}
    \bar{H}_{\rm XY}^{\mathrm{CD}(1)} &= \frac{1}{T} \int_{0}^{T}  J( e^{i \Delta t }\hat{\sigma}_{1}^{+}\hat{\sigma}_{2}^{-} +  e^{-i \Delta t }\hat{\sigma}_{1}^{-}\hat{\sigma}_{2}^{+}) \mathrm{d}t \notag\\
      &= \frac{1}{T}  \frac{J}{i\Delta} \left(e^{i \Delta t }\hat{\sigma}_{1}^{+}\hat{\sigma}_{2}^{-} - e^{-i \Delta t }\hat{\sigma}_{1}^{-}\hat{\sigma}_{2}^{+} \right) \Big|^{T}_{0}.   
      \label{eqn:HCD1}
\end{align}
Because of this periodic structure, the oscillatory contribution can be averaged out by aligning operations at specific time points. To achieve first-order suppression, we choose a matched gate time $T_M$ to satisfy $T_{M}= 2m T_{\Delta}$, $T_{\Delta}\equiv \pi/\abs{\Delta}$, $m \in \mathbb{Z}^{+}$ to make the leading-order error term vanish, $\bar{H}_{\rm XY}^{\mathrm{CD}(1)}=0$. In the following analysis, we focus on the shortest matched gate time ($m=1$) for concreteness, while the cancellation condition holds for arbitrary $m \in \mathbb{Z}^{+}$.

To quantify the performance of gate operations, we use the following fidelity metric defined as
\begin{align}
    F = \abs{\frac{\mathrm{Tr}\left[\hat{U}^{\dagger}_{\mathrm{gate}}(T,0)\hat{U}_{\mathrm{ideal}}(T,0)\right]}{\mathrm{Tr}\left[\hat{U}^{\dagger}_{\mathrm{ideal}}(T,0)\hat{U}_{\mathrm{ideal}}(T,0)\right]}},
    \label{eqn:fidelity}
\end{align}
where $\hat{U}_{\rm ideal}(T,0)$ represents the target evolution of the intended gate operation. This fidelity $F$ quantifies the overlap between the actual evolution $\hat{U}_{\rm gate}(T,0)$ and the ideal evolution $\hat{U}_{\rm ideal}(T,0)$ over the gate duration $T$, while the infidelity $1-F$ characterizes deviations arising from unwanted interactions or imperfections.

Take the idle case as an example, where no active control is applied. The idle gate fidelity under crosstalk dynamics is defined as
\begin{align}
    F^{\rm CD}_{I_{1}I_{2}} = \abs{\frac{\mathrm{Tr}\left[\hat{U}^{\rm{CD} \dagger}_{\rm{idle}}(T,0)\hat{U}_{I_{1}I_{2}}(T,0)\right]}{\mathrm{Tr}\left[\hat{U}^{\dagger}_{I_{1}I_{2}}(T,0)\hat{U}_{I_{1}I_{2}}(T,0)\right]}},
    \label{eqn:FCDidle}
\end{align}
where $\hat{U}^{\rm CD}_{\rm idle}(T,0)$ denotes the evolution in the presence of residual XY coupling, and $\hat{U}_{I_{1}I_{2}}(T,0) = \hat{I}_1 \otimes \hat{I}_2$ represents the ideal idle operation where both qubits stay unchanged.

\begin{figure}[htbp]
\begin{center}
\includegraphics[width=0.6\linewidth]{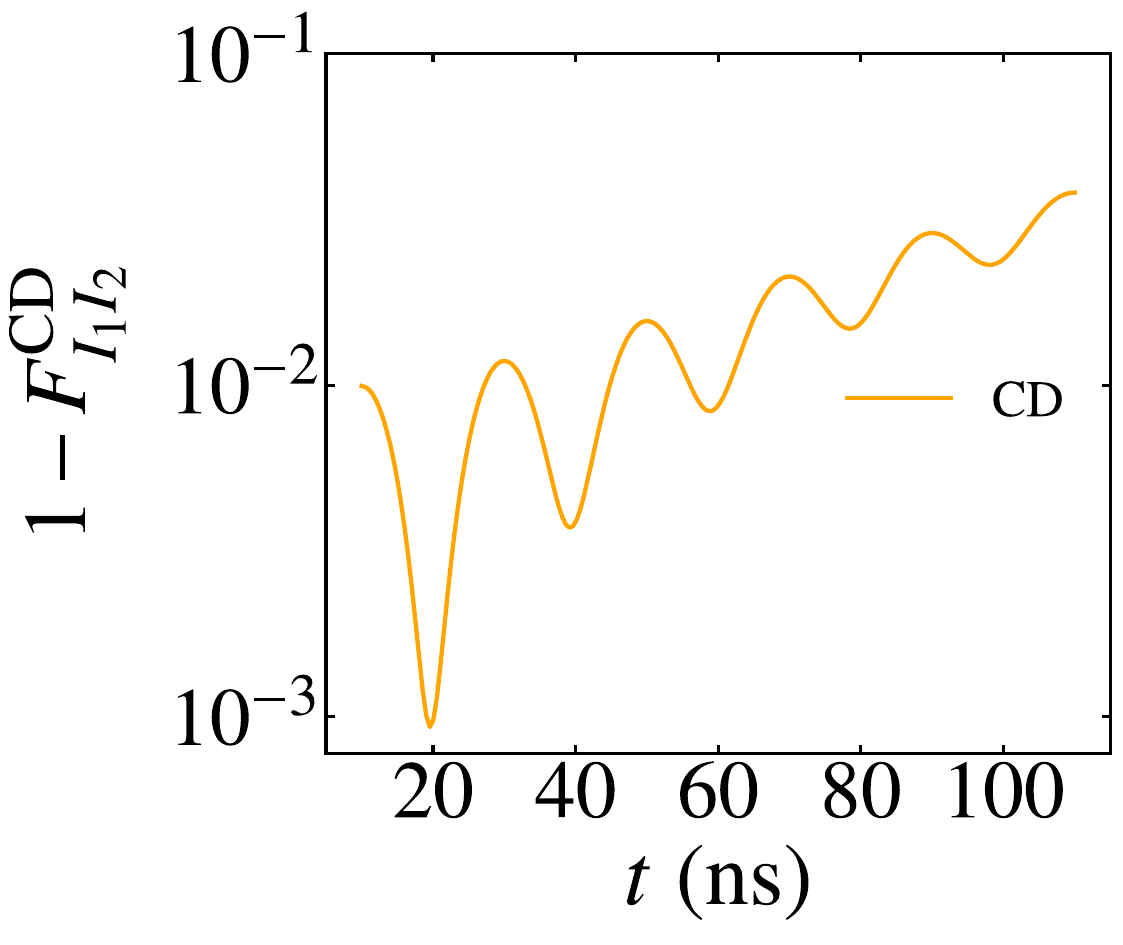}
\caption{Idle gate infidelity, $1 - F^{\rm CD}_{I_1 I_2}$, as a function of time under crosstalk dynamics. Parameters are chosen as $\Delta/2\pi = 50~\mathrm{MHz}$ and $J/2\pi = 5~\mathrm{MHz}$.}
\label{fig:idleIF}
\end{center}
\end{figure}

To examine the gate-time dependence of crosstalk errors, we numerically evaluate the idle gate infidelity for a fixed detuning $\Delta/2\pi = 50~\mathrm{MHz}$, which is used in all numerical simulations. As shown in the simulated infidelity curve in Fig.~\ref{fig:idleIF}, local minima of infidelity appear at even multiples of $T_{\Delta}=10~\mathrm{ns}$, while local maxima occur at odd multiples of $T_{\Delta}$. Therefore, first-order suppression of the XY interaction can be achieved when the gate duration is properly synchronized with the detuning period, without requiring additional control fields.

To further reduce the influence of $\tilde{H}_{\rm XY}(t)$, we introduce active control mechanisms in the following sections. If the qubit frequencies $\omega_{1,2}$ vary with time, the instantaneous detuning $\Delta + \delta(t) = \omega_{1}(t) - \omega_{2}(t)$ generates a time-dependent phase modulation that governs the excitation exchange. By intentionally engineering this detuning through $Z$-control pulses, one can tailor the accumulated phase to minimize the effect of the XY interaction over the gate duration, thereby suppressing crosstalk errors without altering the underlying hardware coupling. In the following, we extend this phase-accumulation mechanism to two active control schemes, frequency modulation and dynamical decoupling, to further suppress higher-order error contributions.

\section{Frequency Modulation}
We present an enhanced control method, termed frequency modulation (FM), which suppresses higher-order gate errors by continuously modulating the qubit frequency via an additional $Z$-control. For simplicity, we first consider the case where the drive is applied to $Q_1$ to perform single-qubit operations, while the $Z$-modulation is applied to the neighboring qubit $Q_2$. This modulation applied to $Q_2$ is in the form of a sinusoidal flux control,
\begin{align}
\hat{H}^{\rm FM}_{Z_2}(t)=\gamma \sin(\tfrac{2\pi N}{T}t)\hat{\sigma}^{z}_{2},
    \label{eqn:HZ2FM}
\end{align}
where $\gamma$ denotes the modulation amplitude, $T$ represents the total gate time, and $N$ specifies the number of modulation cycles executed within the gate duration, such that the modulation period is defined as $\tau = T/N$. This $Z$-drive induces a time-dependent frequency shift on $Q_2$, generating a periodic phase modulation that effectively averages out the crosstalk interaction.

The total Hamiltonian with frequency modulation is
\begin{align}
    \hat{H}^{\rm FM}(t) =& \hat{H}_0 + \hat{H}^{\rm FM}_{Z_2}(t)+ \hat{H}_{\rm XY} + \hat{H}_{\rm drive}(t)
    \notag\\
    =&-\frac{\omega_1}{2} \hat{\sigma}^{z}_1+  \left[\gamma \sin \left(\tfrac{2\pi N}{T}t \right)-\frac{\omega_2}{2} \right] \hat{\sigma}^{z}_2 \notag\\
    &+ J \left( \hat{\sigma}_1^{+} \hat{\sigma}_2^{-} + \hat{\sigma}_1^{-} \hat{\sigma}_2^{+} \right) + \hat{H}_{\rm drive}(t).
\label{eqn:HFM}
\end{align}
To analyze the effect of this modulation, we move to a new rotating frame including the static and the instantaneous frequency of the qubits, referred to as the modulated frame.
The transformation is given by
\begin{align} 
    \tilde{H}^{V}(t) = \hat{V}^{\dagger}(t)\hat{H}(t)\hat{V}(t) + i\Dot{\hat{V}}^{\dagger}(t)\hat{V}(t),
    \label{eqn:HVTransform}
\end{align}
with the unitary transformation defined as:
\begin{align}
    \hat{V}(t) = \exp \left\{-i\left( -\frac{\omega_{1}}{2}t\hat{\sigma}_{1}^z +  \left[\tfrac{\gamma T}{\pi N} \sin^{2}{(\tfrac{\pi N}{T}t)}-\frac{\omega_2}{2} t\right] \hat{\sigma}_{2}^z\right) \right\}.
    \label{eqn:UTransform}
\end{align}
The XY interaction in this  modulated frame becomes
\begin{align}
     \tilde{H}_{\rm XY}^{V}(t) = J \left( e^{i \left[\Delta t +2\alpha(t)\right]}\hat{\sigma}_{1}^{+}\hat{\sigma}_{2}^{-} +  e^{-i \left[\Delta t +2\alpha(t)\right]}\hat{\sigma}_{1}^{-}\hat{\sigma}_{2}^{+} \right),
     \label{eqn:XYVframe}
\end{align}
where $\alpha(t) = \frac{\gamma T}{\pi N} \sin^{2}{(\frac{\pi N}{T}t)}$ corresponds to a time-dependent modulation phase. The drive term remains identical to that in the static rotating frame, $\tilde{H}^{V}_{\rm drive}(t)=\tilde{H}_{\rm drive}(t)$, since it acts only on $Q_1$ and is therefore unaffected by the modulation on $Q_2$.

The gate unitary under frequency modulation is then given by
\begin{align}
     \hat{U}^{\rm FM}_{\rm gate}(T,0) =  \mathcal{T}\exp \left\{ -i\int_{0}^{T} \left[\tilde{H}_{\rm drive}(t)+ \tilde{H}^{V}_{\rm XY}(t) \right] \mathrm{d}t \right\}.
\label{eqn:UFG}
\end{align}
The frequency-modulated frame coincides with the static rotating (operation) frame at the beginning and end of the gate, since the additional rotation generated by $\hat{H}^{\mathrm{FM}}_{Z_2}(t)$ vanishes at $t=0$ and $t=T$. Hence, the gate evolution is identical in both frames, while the modulation only modifies the intermediate dynamics.

To analyze the effect of crosstalk under frequency modulation, we expand the gate unitary using the Magnus expansion
\begin{align}
\hat{U}^{\rm FM}_{\rm gate}(T,0) &= e^{-iT \left[\bar{H}^{\mathrm{FM}(1)}_{\mathrm{gate}}+ \bar{H}^{\mathrm{FM}(2)}_{\mathrm{gate}}+\cdots \right] },
\label{eqn:UFMmagnus}
\end{align}
with the first-order contribution from the XY interaction given by
\begin{align}
    \bar{H}^{\mathrm{FM}(1)}_{\rm XY}&= \frac{1}{T} \int_{0}^{T}  \tilde{H}^V_{\mathrm{XY}} (t)\mathrm{d}t.
\end{align}
We define the first-order crosstalk error as the sum of the magnitudes of the coefficients preceding the first-order operators,
\begin{align}
    \varepsilon^{\mathrm{FM}(1)}\equiv \left|\frac{J}{T} \int_{0}^{T}  e^{i\left[\Delta t + 2\alpha(t)  \right]}\ \mathrm{d}t \right| +\left|\frac{J}{T} \int_{0}^{T}  e^{-i \left[ \Delta t + 2\alpha(t)  \right]}\ \mathrm{d}t \right|.
    \label{eqn:eFM1}
\end{align}

For a given gate time $T$ and detuning $\Delta$, we can find the optimal modulation amplitude $\gamma^{N}_{\rm opt}$ at the assigned modulated cycle number $N$ to minimize the first-order crosstalk error $\varepsilon^{\mathrm{FM}(1)}$.  When choosing a matched gate time $T_M= 2m\pi/|\Delta|$, $m \in \mathbb{Z}^{+}$, the first-order error averages out to zero as in the crosstalk dynamics case, [Eq.~\eqref{eqn:HCD1}]. Frequency modulation can further be used to suppress the second-order crosstalk error, which depends on the intended operation through $\hat{H}_{\rm drive}(t)$.

We will show how frequency modulation suppresses second-order crosstalk errors and improves gate fidelity using the idle gate and the single-qubit $X_1$ gate as representative examples. While the present analysis assigns the gate operation to $Q_1$ and the modulation to its neighboring qubit $Q_2$, similar control strategy can be applied with both the drive and the modulation acting on a single qubit. Detailed derivations and extended analysis, covering the first-order optimization for nonmatched gate times, the single-qubit scheme of drive and modulation, and the case of parallel $X_1X_2$ gate operations, are provided in Appendix~\ref{App:FMdetails}.

\subsection{Idle gate under frequency modulation \label{subsec:FMIdle}}

We consider the idle gate operation with target evolution $\hat{U}_{I_{1}I_{2}}(T,0) = \hat{I}_1 \otimes \hat{I}_2$ under no drive, $\hat{H}_{\rm drive}(t)=0$.
The second-order error during idle gate is quantified as
\begin{align}
     &\varepsilon^{\rm{FM}(2)}_{\rm{idle}}=\frac{J^2}{T} \abs{\int_{0}^{T} \mathrm{d} t_{1} \int _{0}^{t_1} \mathrm{d} t_{2} \, E^{(2)}_{\mathrm{idle}}(t_1,t_2)},\notag\\
     &E^{(2)}_{\mathrm{idle}}(t_1,t_2)=\sin\left(\Delta (t_1 - t_2)+2 \left[\alpha(t_1) - \alpha(t_2)\right] \right).
\end{align}
For $\Delta/2\pi=50$~MHz and a matched gate time $T_M=2\pi/\abs{\Delta}=20$~ns, we obtain the optimized modulation amplitudes $\gamma^{N}_{\mathrm{opt}}$ that minimize the second-order crosstalk error $\varepsilon^{\mathrm{FM}(2)}_{\mathrm{idle}}$ for each modulation cycle number $N$ and simulate the gate dynamics to evaluate the resulting suppression of crosstalk errors. Unless otherwise stated, the numerical results presented in the main text are obtained using this choice of detuning and matched gate time. Details of the second-order error derivation and optimization procedure are provided in Appendix~\ref{App:FMIdle}.

\begin{figure}[htbp]
\begin{center}
\includegraphics[width=1\linewidth]{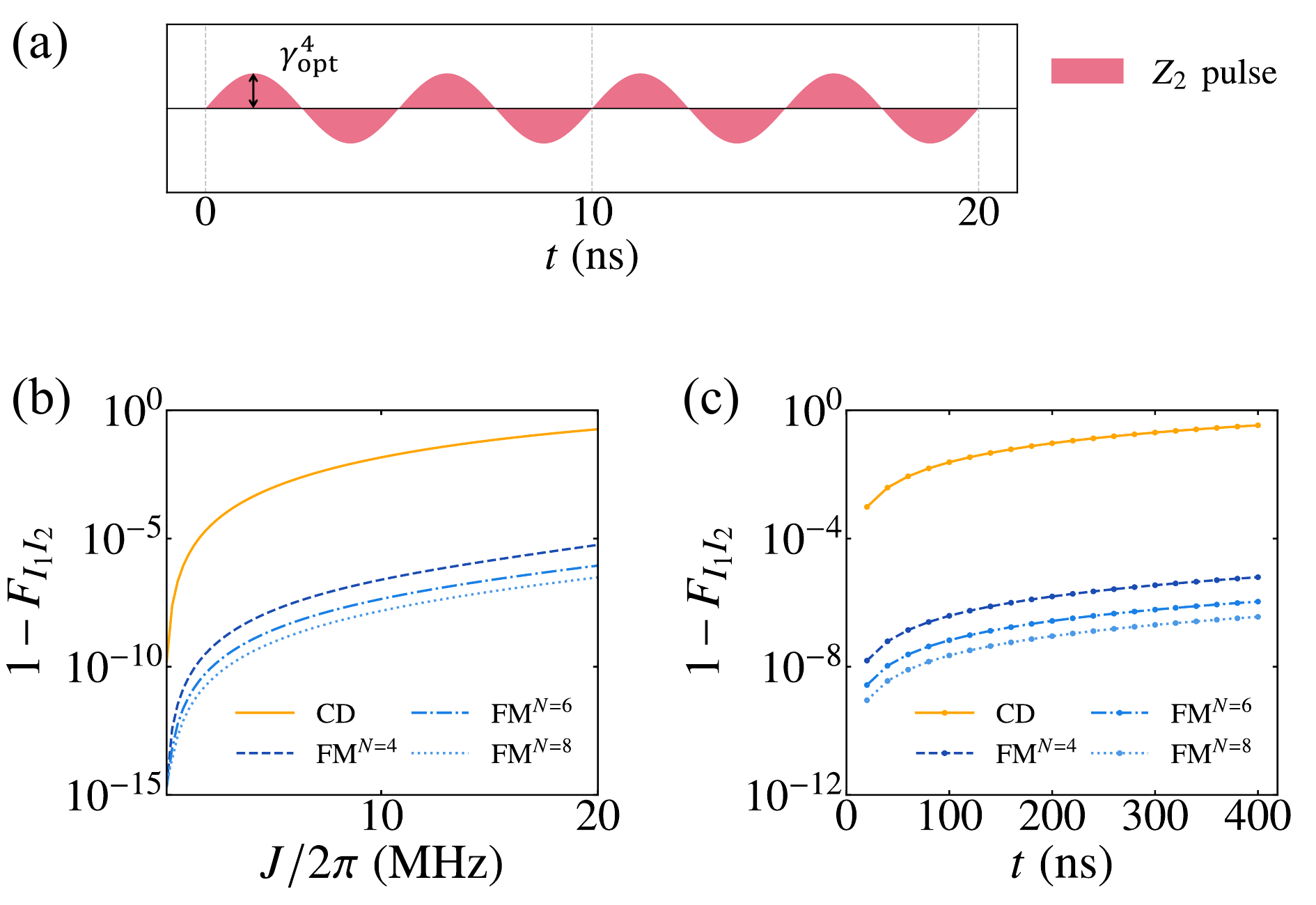}
\caption{Suppression of idle gate infidelity under frequency modulation.
(a) Waveform of a sinusoidal $Z_2$ drive with $N=4$ modulation cycles applied to $Q_2$.
(b) Idle gate infidelity of one idle operation as a function of the coupling strength $J$ for different numbers of modulation cycles $N$, evaluated at the matched gate time $T_M = 20$~ns.
(c) Idle gate infidelity of consecutive idle operations as a function of time for different numbers of modulation cycles $N$, with $T_M = 20$~ns and $J/2\pi = 5$~MHz. Results under crosstalk dynamics (CD) are shown for comparison.
}
\label{fig:FMidleInfidelity}
\end{center}
\end{figure}

To assess the performance of the scheme, we simulate the idle gate with frequency modulation at the optimized modulation amplitude $\gamma^{N}_{\mathrm{opt}}$ for each $N$. An example of the applied modulation, corresponding to the $N=4$ case, is illustrated in Fig.~\ref{fig:FMidleInfidelity}(a).  We first examine how the infidelity depends on the crosstalk coupling strength $J$. As shown in Fig.~\ref{fig:FMidleInfidelity}(b), frequency modulation (FM) largely suppresses the idle gate infidelity, yielding an improvement of more than four orders of magnitude compared with the crosstalk-dynamics (CD) case. The suppression becomes even stronger with increasing modulation cycles, reaching approximately six orders of magnitude at $N=8$, reflecting more efficient averaging of the XY crosstalk, albeit at the expense of a stronger modulation amplitude.

We further simulate a sequence of consecutive idle gates to evaluate the long-term stability of the suppression. As shown in Fig.~\ref{fig:FMidleInfidelity}(c), even after 20 consecutive operations, the idle gate infidelity shows an improvement of more than four orders of magnitude with FM compared with the unmodulated case, indicating consistent suppression of the XY crosstalk.

\subsection{$X_1$ gate under frequency modulation \label{subsec:FMX1}}
We next consider the single-qubit $X_1$ gate, whose target operation is $\hat{U}_{X_{1}I_{2}}(T,0)={-i}\hat{\sigma}_1^x\otimes\hat{I}_2$. The gate is generated by the driving Hamiltonian $\tilde{H}^{V}_{\mathrm{drive}}(t) = \Omega_{1}(t)\hat{\sigma}_{1}^{x}$ 
with a smoothly varying envelope $ \Omega_{1}(t) =\Omega_{1x}\sin(\frac{\pi}{T}t)$, $\Omega_{1x}=\frac{\pi^2}{4T}$.  The pulse satisfies $\int_{0}^{T}\Omega_{1}(t) \mathrm{d}t=\frac{\pi}{2}$, yielding a total $\pi$ rotation about the $x$-axis on the Bloch sphere of $Q_1$.

\begin{figure}[htbp]
\begin{center}
\includegraphics[width=1\linewidth]{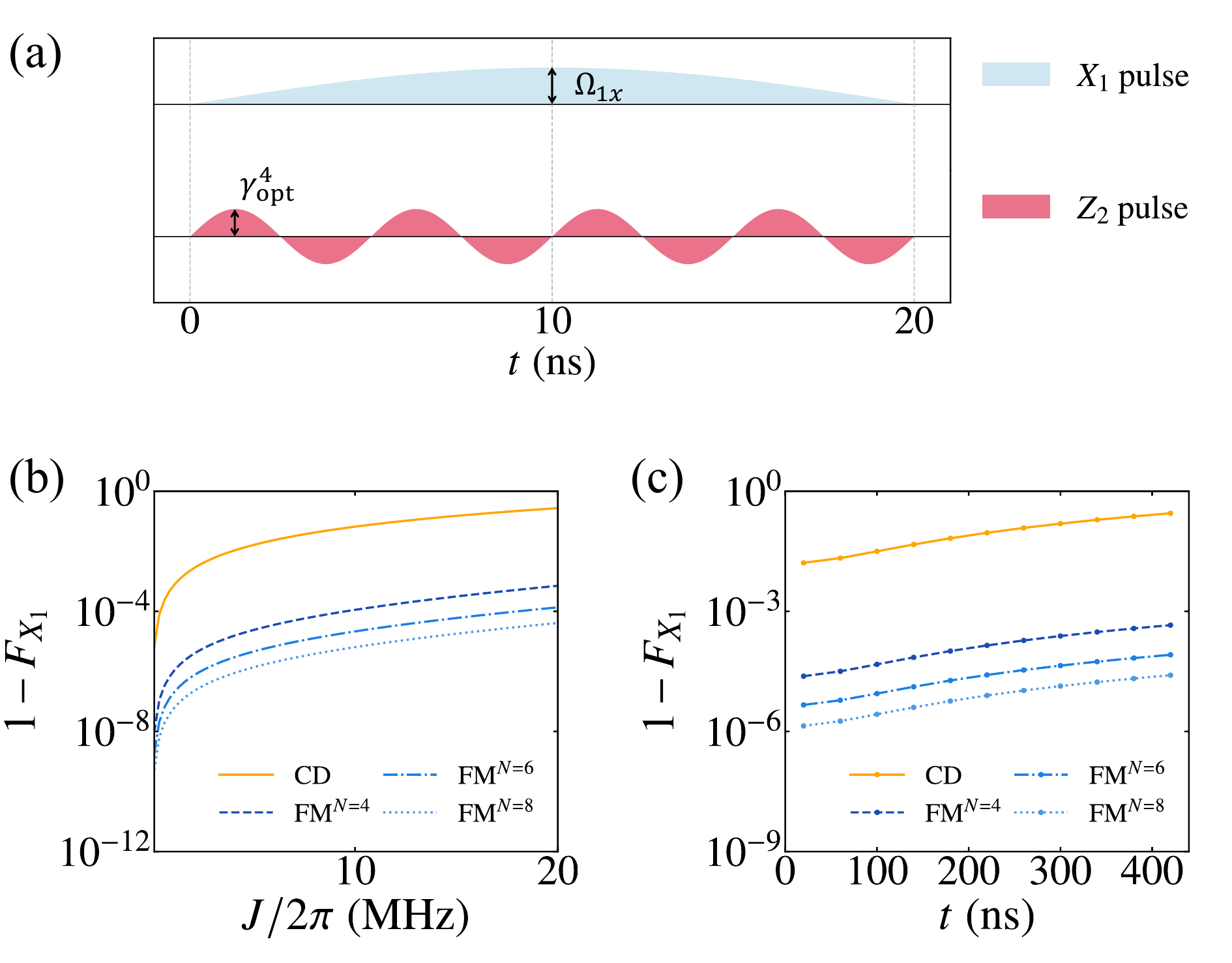} 
\caption{Suppression of $X_{1}$ gate infidelity under frequency modulation. (a) Waveforms of the $X_1$ control pulses applied to $Q_1$ and the sinusoidal $Z_2$ drive with $N=4$ modulation cycles applied to $Q_2$.
(b) $X_{1}$ gate infidelity of one $X_1$ operation as a function of the coupling strength $J$ for different numbers of modulation cycles $N$.
(c) $X_{1}$ gate infidelity of consecutive $X_1$ operations as a function of time for different numbers of modulation cycles $N$, with a fixed $J/2\pi = 5$~MHz. Odd numbers of consecutive $X_{1}$ gates are applied to perform a net $X_1$ operation.}
    \label{fig:FMX1Infidelity}
\end{center}
\end{figure}

The second-order error is given by
\begin{align}
     &\varepsilon^{\mathrm{FM}(2)}_{X_1} = 2 \abs{ \frac{iJ}{2T} \int _{0}^{T} \mathrm{d}t_{1} \int_{0}^{t_1} \mathrm{d}t_{2}\, E^{(2)}_{X_1}(t_1,t_2)} +\varepsilon^{\mathrm{FM}(2)}_{\mathrm{idle}},\notag\\
     &E^{(2)}_{X_1}(t_1,t_2) =\Omega_{1}(t_{1})e^{i\left[\Delta t_2 + 2\alpha(t_2)\right]}-\Omega_{1}(t_2)e^{i\left[\Delta t_{1} + 2\alpha(t_1)\right]}.
\end{align}
For $\Delta/2\pi=50$~MHz and a matched gate time $T_M=20$~ns, we obtain the optimized modulation amplitudes $\gamma^{N}_{\mathrm{opt}}$ that minimize the second-order crosstalk error $\varepsilon^{\mathrm{FM}(2)}_{X_1}$ for each modulation cycle number $N$ (see Appendix~\ref{App:FMX1} for details).
Using these optimized values, we simulate the gate dynamics to evaluate the resulting improvement in fidelity, with Fig.~\ref{fig:FMX1Infidelity}(a) providing a concrete example of the frequency modulation and control drive waveforms for the $N=4$ case.

As shown in Fig.~\ref{fig:FMX1Infidelity}(b), FM reduces the $X_1$ gate infidelity by more than two orders of magnitude compared with the crosstalk-dynamics case, with further improvement as the number of modulation cycles $N$ increases. To assess performance under extended operations, we simulate sequences of consecutive $X_1$ gates. Figure~\ref{fig:FMX1Infidelity}(c) shows that even after 21 consecutive gates, the infidelity remains more than two orders of magnitude lower than in the absence of FM, illustrating consistent suppression of XY crosstalk during repeated single-qubit operations.

These results show that frequency modulation effectively suppresses crosstalk errors in both idle and active single-qubit $X$ operations, providing enhanced single-qubit control. Since the modulation parameters are optimized for a given detuning, nearby qubits with similar frequency offsets can be simultaneously protected, enabling scalable suppression of XY crosstalk, as further analyzed in a later section. The same principle also applies when both the drive and modulation are applied to the same qubit, where the drive term is modified accordingly (Appendix~\ref{app:SingleQFM}), and can be further extended to parallel single-qubit operations ($X_1X_2$) (Appendix~\ref{app:X1X2FM}).

\section{Dynamical Decoupling}
Dynamical decoupling (DD) is a widely established technique for suppressing unwanted couplings in quantum systems through carefully designed pulse sequences~\cite{Viola1998,Viola1999,Ng2011,Pokharel2018,Ezzell2023,Tripathi2025}. By averaging out residual interaction terms over the evolution period, DD effectively mitigates coherent crosstalk. In this section, we apply DD strategies specifically to suppress errors arising from residual XY interactions.

To suppress the XY interaction, we apply $Z_{2}$ gates to $Q_{2}$ at regular intervals $\tau = T/S$, where $S$ is the number of segments chosen to be an even integer and the total duration is set to the matched gate time $T = T_{M} = 2\pi/\abs{\Delta}$. We model each $Z_{2}$ gate as a finite-width sinusoidal flux pulse centered at $t=s\tau$,
\begin{align}
    \hat{H}^{\rm DD}_{Z_2}(t) = \sum_{s=1}^{S}\frac{\pi}{2} \delta_{w}(t-s\tau)\hat{\sigma}^{z}_{2},
     \label{eqn:HZ2DD}
\end{align}
where $\delta_{w}(t)$ is a nascent delta function of width $w$,
\begin{align}
\delta_w(t) \equiv
\begin{cases}
\dfrac{\pi}{2w}\cos\left(\dfrac{\pi t}{w}\right), & |t|\le w/2, \\[6pt]
0, & \text{otherwise},
\end{cases}
\label{eqn:deltaw}
\end{align}
which approaches the Dirac delta function in the zero-width limit,
$\lim_{w \to 0} \delta_w(t) = \delta(t)$. Each $Z_{2}$ pulse imparts a $\pi$ phase shift to $Q_{2}$, periodically inverting the XY interaction and effectively averaging out the crosstalk.

The total Hamiltonian under dynamical decoupling is given by
\begin{align}
    \hat{H}^{\rm DD}(t) =& \hat{H}_0 + \hat{H}^{\rm DD}_{Z_2}(t)+ \hat{H}_{\rm XY} + \hat{H}_{\rm drive}(t)
    \notag\\
    =&-\frac{\omega_1}{2} \hat{\sigma}^{z}_1+  \left[\sum_{s=1}^{S}\frac{\pi}{2}\delta_w(t-s\tau)-\frac{\omega_2}{2} \right] \hat{\sigma}^{z}_2 \notag\\
    &+ J \left( \hat{\sigma}_1^{+} \hat{\sigma}_2^{-} + \hat{\sigma}_1^{-} \hat{\sigma}_2^{+} \right) + \hat{H}_{\rm drive}(t).
\label{eqn:HDD}
\end{align}
To analyze the effect of this dynamical decoupling, we move to the operation frame rotated by the static qubit frequencies as defined by Eqs.~\eqref{eqn:rotQubit} and ~\eqref{eqn:HVCD}.

The Hamiltonian in the operation frame takes the form
\begin{align}
    \tilde{H}^{\rm DD}(t) =& \tilde{H}^{\rm DD}_{Z_2}(t)+ \tilde{H}_{\rm XY}(t) + \tilde{H}_{\rm drive}(t).
\end{align}
In the limit of short pulses, $w \rightarrow 0$, we approximate all $Z_{2}$ operations as instantaneous. The overall gate unitary under dynamical decoupling is then obtained by inserting a $Z_{2}$ gate into the crosstalk evolution at the end of each time segment,
\begin{align}
    \hat{U}^{\mathrm{DD}}_{\mathrm{gate}}(T,0)
    = \hat{Z}_{2}\hat{U}^{\mathrm{CD}}[S\tau,(S-1)\tau]\cdots \hat{Z}_{2}\hat{U}^{\mathrm{CD}}[\tau,0],
    \label{eqn:UDDgate}
\end{align}
where the crosstalk evolution is
\begin{align}
\hat{U}^{\mathrm{CD}}(t_{2},t_{1})
= \mathcal{T}\exp\!\left[-i\!\int_{t_{1}}^{t_{2}}
\left[\tilde{H}_{\rm XY}(t)+\tilde{H}_{\rm drive}(t)\right]\,dt\right].
    \label{eqn:UCD}
\end{align}

We can absorb the instantaneous $Z_{2}$ gates into the even segments, such that the segment evolution operators take the form
\begin{align}
\hat{U}_{s} &=
\begin{cases}
\hat{U}^{\mathrm{CD}}[s\tau,(s-1)\tau], 
& s \ \text{odd}, \\[6pt]
\hat{Z}_{2}\,\hat{U}^{\mathrm{CD}}[s\tau,(s-1)\tau]\,\hat{Z}_{2}, 
& s \ \text{even},
\end{cases}
\label{eqn:Us}
\end{align}
and the total DD evolution becomes
\begin{align}
\hat{U}^{\mathrm{DD}}_{\mathrm{gate}}(T,0)
= \prod_{s=1}^{S}\hat{U}_{s},
\label{eqn:UDDgateproduct}
\end{align}
where the product is time ordered with earlier segments appearing on the right.

We define the segment Hamiltonian $\tilde{H}_{s}(t)$ through $\hat{U}_{s}  =\mathcal{T}\exp\!\left[-i \int_{(s-1)\tau}^{s\tau} \tilde{H}_{s}(t) \, \mathrm{d}t \right]$. Using the identity $U e^{M}U^{\dagger} = e^{U M U^{\dagger}}$, we have
\begin{align}
\tilde{H}_{s}(t) &=
\begin{cases}
\tilde{H}_{\rm drive}(t)+ \tilde{H}_{\rm XY}(t), 
& s \ \text{odd}, \\[6pt]
\hat{Z}_{2}\left[\tilde{H}_{\rm drive}(t)+ \tilde{H}_{\rm XY}(t) \right]\hat{Z}_{2}, 
& s \ \text{even}.
\end{cases}
\label{eqn:Hs}
\end{align}

In even segments, the XY interaction acquires a minus sign because $\hat{Z}_{2}$ anticommutes with $\hat{\sigma}_{2}^{\pm}$. Drive components involving $\hat{\sigma}_{2}^{x}$ or $\hat{\sigma}_{2}^{y}$ would also flip sign under the same transformation. The resulting Hamiltonian can be written in an effective form that incorporates the alternating sign induced by the DD sequence,
\begin{align}
\tilde{H}^{\rm DD}_{\rm eff}(t)=
f(t)\left[\tilde{H}_{\mathrm{XY}}(t) + \tilde{H}^{Q_2}_{\rm drive}(t)\right]+\tilde{H}^{Q_1}_{\rm drive}(t),
\label{eqn:HDDeff}
\end{align}
where we have introduced a sign function $f(t)=(-1)^{s-1}$ for $t\in[(s-1)\tau,s\tau]$. Here, $\tilde{H}^{Q_1}_{\mathrm{drive}}(t)$ denotes the transverse (XY) control drive term acting on $Q_1$, while $\tilde{H}^{Q_2}_{\mathrm{drive}}(t)$ represents the transverse control drive term acting on $Q_2$, which acquires the sign-flipping factor $f(t)$ due to the applied DD sequence. The alternating sign leads to an effective averaging of the XY interaction, forming the basis for crosstalk suppression achieved by dynamical decoupling.

Using the Magnus expansion, the gate evolution under dynamical decoupling can be expressed as 
\begin{align}
   \hat{U}^{\mathrm{DD}}_{\mathrm{gate}}(T,0)
   = e^{-iT\left[\bar{H}^{\mathrm{DD}(1)}_{\mathrm{gate}}
   +\bar{H}^{\mathrm{DD}(2)}_{\mathrm{gate}}+\cdots\right]},
    \label{eqn:magnus_DD}
\end{align}
with the first-order contribution from the XY interaction given by 
\begin{align}
    \bar{H}^{\mathrm{DD}(1)}_{\mathrm{XY}}
    = \frac{1}{T}\sum_{s=1}^{S}\int_{(s-1)\tau}^{s\tau}
      (-1)^{s-1}\tilde{H}_{\mathrm{XY}}(t)\,\mathrm{d}t,
    \label{eqn:general1stDD}
\end{align}
which is the sum of integrals over the XY crosstalk Hamiltonian in each segment.

The corresponding first-order crosstalk error is (see Appendix~\ref{App:DD1sterror})
\begin{align}
    \varepsilon^{\mathrm{DD}(1)}
    \equiv 2 \abs{\frac{J}{T}
    \sum_{s=1}^{S}\frac{(-1)^{\,s-1}}{i\Delta}
    \left[e^{i\Delta s\tau}-e^{i\Delta(s-1)\tau}\right]}.
    \label{eqn:generalDDerror}
\end{align}
The factor $(-1)^{s-1}$ reflects the sign inversion of $\tilde{H}_{\mathrm{XY}}(t)$ in every even segment, as seen directly from Eq.~\eqref{eqn:general1stDD}. Due to the periodicity of $e^{\pm i\Delta t}$, the above expression vanishes for any even number of segments $S \ge 4$, reproducing the same first-order cancellation that occurs at the matched gate time.

We will use the idle gate and the single-qubit $X_{1}$ gate as representative examples to illustrate how dynamical decoupling suppresses second-order XY crosstalk and improves gate fidelity. In the main analysis, the control drive is applied to $Q_{1}$ while the DD $Z_{2}$ pulses are applied to its neighbor $Q_{2}$. An equivalent single-site protocol, in which both the drive and the decoupling pulses act on the same qubit, is also possible. Detailed derivations and further extensions, including the single-site version and parallel $X_{1}X_{2}$ operations, are provided in Appendix~\ref{App:DDdetails}.

\subsection{Idle gate under dynamical decoupling \label{subsec:DDIdle}}
We consider the specific case of $S=4$ segments, corresponding to a DD sequence consisting of four equally spaced $Z_{2}$ pulses over the full gate duration $T_M$, with a segment interval $\tau = T_M/4$. This sequence will be referred to as DD Z-4. The second-order errors of the idle gate under crosstalk dynamics and under dynamical decoupling are computed as (see Appendix~\ref{App:DDidle})
\begin{align}
    & \varepsilon^{\mathrm{CD}(2)}_{\mathrm{idle}}= 2\left| \frac{J^2 }{2\Delta} \right| , \notag \\ & \varepsilon^{\mathrm{DD}(2)}_{\mathrm{idle}} = 2\left|\frac{\pi-4}{2\pi}\frac{J^2 }{\Delta}\right|.
\end{align}
Since $\left|\frac{\pi-4}{\pi}\right| < 1 $, the second-order error is reduced under
DD, which is expected to improve the resulting gate fidelity.

\begin{figure}[htbp]
\begin{center}
\includegraphics[width=1\linewidth]{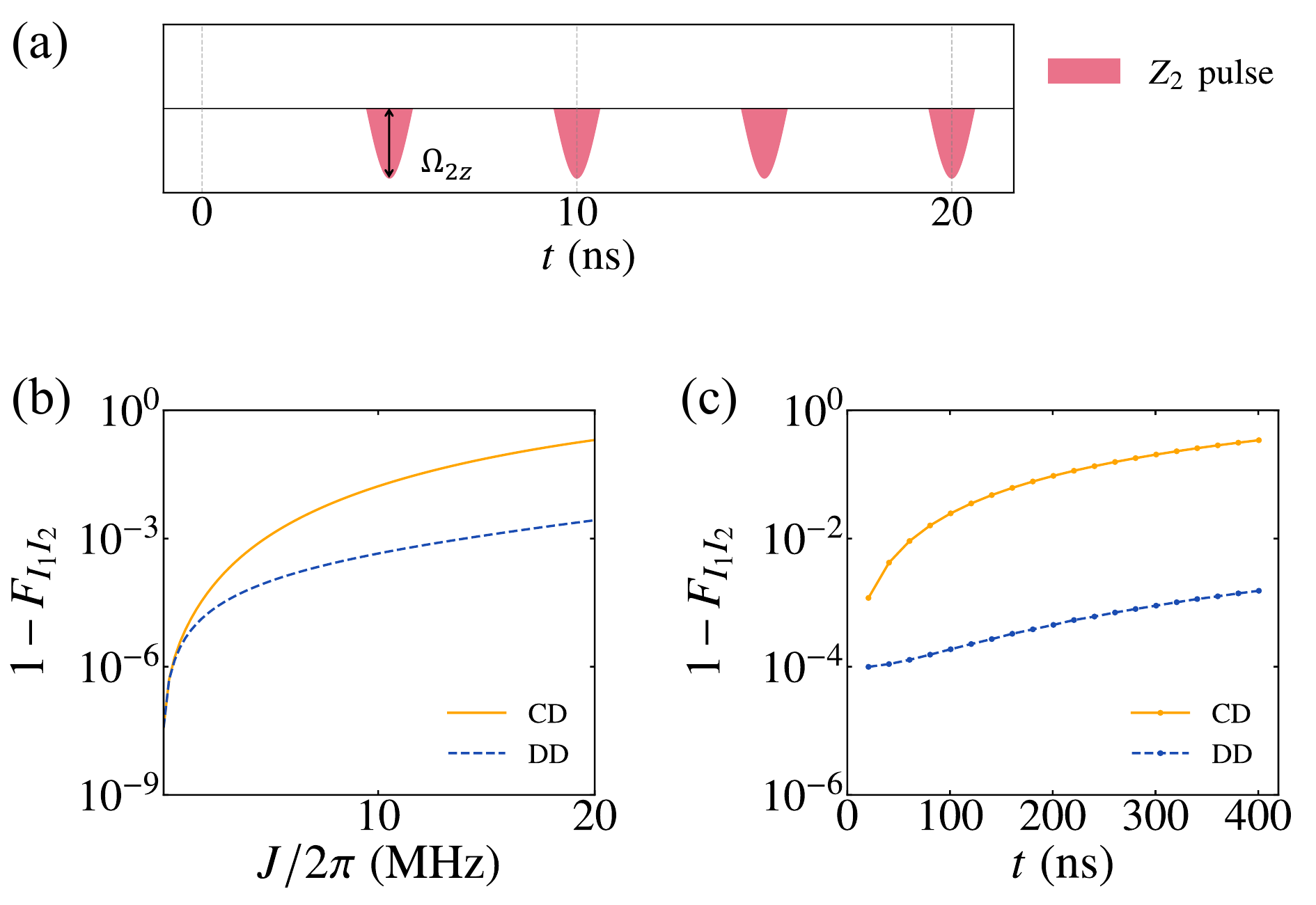}
\caption{Suppression of idle gate infidelity under dynamical decoupling.
(a) Waveform of the DD Z-4 sequence containing four $Z_2$ gates applied to $Q_2$.
(b) Idle gate infidelity of one idle operation as a function of the coupling strength $J$ under DD Z-4.
(c) Idle gate infidelity of consecutive idle operations as a function of time under DD Z-4, with $J/2\pi = 5$~MHz.}
\label{fig:DDidleInfidelity}
\end{center}
\end{figure}

While the analytical calculations above assume instantaneous $Z_{2}$ operations, in numerical simulations we include their finite duration to reflect realistic pulse shapes. For the DD Z-4 sequence with $\tau = T_{M}/4$, we model each $Z_{2}$ pulse with a finite width $w = \tau/4$. The total evolution time therefore includes this pulse duration, and for sequences of consecutive idle gates the fidelity is evaluated at $nT_M + \frac{w}{2}$, where $n$ is the number of idle gates. The $Z_{2}$ pulses lower the frequency of $Q_{2}$, increasing its detuning from $Q_{1}$ for better suppression of the XY interaction. The corresponding DD Z-4 pulse waveform for a single idle gate is shown in Fig.~\ref{fig:DDidleInfidelity}(a).

As shown in Fig.~\ref{fig:DDidleInfidelity}(b), for a detuning of $\Delta/2\pi = 50$~MHz and the matched gate time $T_M = 20$~ns, the DD Z-4 sequence reduces the idle gate infidelity by approximately one order of magnitude compared with the crosstalk-dynamics case for $J/2\pi \geq 5$~MHz. To assess performance over extended idle durations, we simulate sequences of consecutive idle gates. Figure~\ref{fig:DDidleInfidelity}(c) shows that even after 20 consecutive operations, the idle gate infidelity with DD is more than two orders of magnitude lower than without DD, indicating stable suppression of the XY interaction over repeated idling. These results show that DD Z-4 suppresses both the idle gate error and its accumulation over repeated operations, enabling more robust idling over extended durations.

\subsection{$X_1$ gate under dynamical decoupling \label{subsec:DDX1}}
To examine the effect of dynamical decoupling on active single-qubit operations, we analyze an $X_{1}$ gate applied within the DD Z-4 sequence, where the matched gate time $T_M$ is divided into four equal segments of duration $\tau = T_M/4$. The gate is decomposed into two $\sqrt{X_1}$ rotations applied in the odd segments between adjacent $Z_{2}$ pulses. For simplicity, we apply the control drive only during odd segments to ensure that the drive term remains unaffected by the $Z_{2}$ pulses in Eq.~\eqref{eqn:Hs}.

\begin{figure}[htbp]
\begin{center}
\includegraphics[width=1\linewidth]{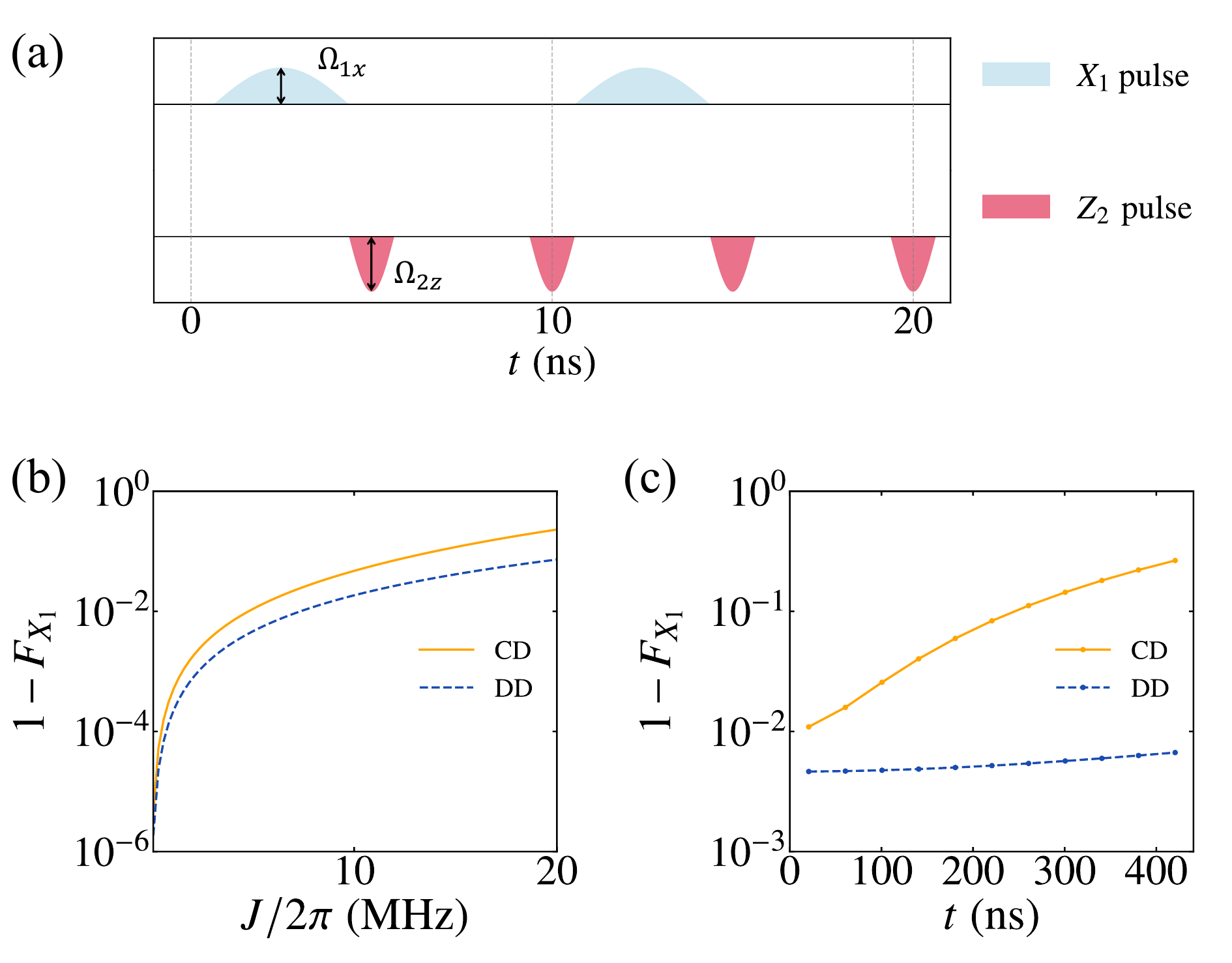}
\caption{Suppression of $X_{1}$ gate infidelity under dynamical decoupling.
(a) Waveforms of the $X_1$ control pulses (generated by two $\sqrt{X_1}$ pulses) applied to $Q_1$ and the DD Z-4 sequence containing four $Z_2$ gates applied to $Q_2$.
(b) $X_{1}$ gate infidelity of one $X_1$ operation as a function of the coupling strength $J$ under DD Z-4.
(c) $X_{1}$ gate infidelity of consecutive $X_1$ operations as a function of time under DD Z-4, with $J/2\pi = 5$~MHz.}
    \label{fig:DDX1Infidelity}
\end{center}
\end{figure}

The corresponding drive Hamiltonian is given by
\begin{align}
&\tilde{H}^{Q_1}_{\mathrm{drive}} (t)
    = \Omega_{1x}(w)
      \cos\left(\frac{\pi}{\tau - w}
      \left[t - \left(s-\tfrac{1}{2}\right)\tau\right]\right)\hat{\sigma}_{1}^{x}, \notag \\
    &t \in [(s-1)\tau+\tfrac{w}{2},\, s\tau-\tfrac{w}{2}],\quad s \in \{1,3\},
    \label{eqn:X_1DDgeneric}
\end{align}
and is equal to zero otherwise.
Here $w$ is the pulse width of the $Z_2$ gate, and the drive amplitude is
$\Omega_{1x}(w) = \frac{\pi^{2}}{8(\tau - w)}$. An example of waveforms with $w=\tau/4$ is illustrated in Fig.~\ref{fig:DDX1Infidelity}(a).

Under the short-pulse limit, $w \rightarrow 0$, the second-order errors with and without dynamical decoupling are (see Appendix~\ref{App:DDX1})
\begin{align}
    &\varepsilon^{\mathrm{CD}(2)}_{X_1}= 2 \left|{ \frac{J^2 }{2\Delta}}\right| + 2 \left|{\frac{J}{4}}\right| , \notag \\ &\varepsilon^{\mathrm{DD}(2)}_{X_1}= 2 \left|{\frac{\pi-4}{2\pi}\frac{J^2 }{\Delta}}\right| +2 \left|{\frac{J}{4} }\right|,
\end{align}
with $\varepsilon^{\mathrm{DD}(2)}_{X_1} < \varepsilon^{\mathrm{CD}(2)}_{X_1}$ indicating suppression of crosstalk errors during the $X_1$ gate under dynamical decoupling.

We then consider finite-width $Z_2$ pulses for numerical simulations of the DD sequence, as shown in Fig.~\ref{fig:DDX1Infidelity}(a). The CD case without $Z_2$ pulses is simulated with $w=0$, consistent with the analytical derivation. Figure~\ref{fig:DDX1Infidelity}(b) shows the resulting $X_{1}$-gate infidelity as a function of the crosstalk coupling strength $J$. The DD Z-4 sequence reduces the infidelity by approximately $0.37$ orders of magnitude for $J/2\pi \geq 5$~MHz, with smaller improvements than in the idle gate case due to the presence of the drive during operation.

We also simulate sequences of consecutive $X_{1}$ gates, as shown in Fig.~\ref{fig:DDX1Infidelity}(c). Without DD, infidelity grows rapidly with the number of gates, while DD Z-4 slows this accumulation. After 21 consecutive gates, the DD-protected sequence achieves more than one order-of-magnitude reduction compared with the unprotected case, indicating sustained suppression of XY crosstalk during repeated single-qubit operations.

\section{Scalable XY-crosstalk suppression}
The proposed XY-crosstalk suppression methods are naturally extensible to multi-qubit settings, enabling scalable qubit architectures. As quantum processors expand to larger systems with denser connectivity, residual XY-type interactions increasingly give rise to coherent crosstalk, posing a growing challenge for high-fidelity control. To assess whether the suppression remains effective in large-scale architectures, we generalize the two-qubit model to a five-qubit configuration and evaluate the corresponding performance.

\begin{figure}[htbp]
\begin{center}
\includegraphics[width=0.5\linewidth]{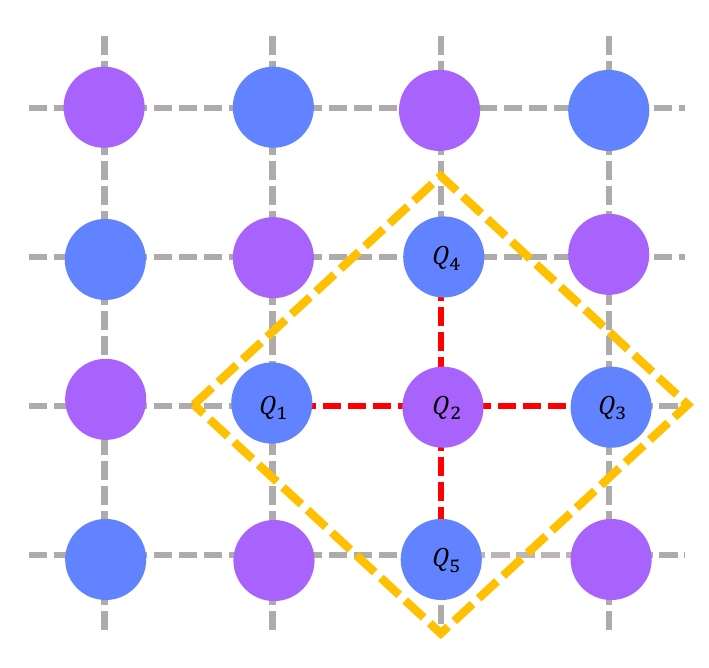} 
\caption{A five-qubit subarray within a nearest-neighbor qubit array is indicated by the dashed orange outline.
The XY interactions involving the central qubit $Q_2$ are highlighted by red dashed lines.}
    \label{fig:5q}
\end{center}
\end{figure}

We study a five-qubit layout in which $Q_2$ couples to its four nearest neighbors (Fig.~\ref{fig:5q}). The Hamiltonian of the five-qubit system is
\begin{align}
    \hat{H}_{\text{5Q}}(t) &=-\sum_{i=1}^{5}\frac{\omega_{i}}{2}\hat{\sigma}_{i}^z + \sum_{\substack{j=1 \\ j \neq 2}}^{5}  J( \hat{\sigma}_{j}^{+}\hat{\sigma}_{2}^{-} +  \hat{\sigma}_{j}^{-}\hat{\sigma}_{2}^{+}).
    \label{eqn:H5Q}
\end{align}
In the operation frame, i.e., the rotating frame at the static qubit frequencies, the five-qubit Hamiltonian becomes
\begin{align}
     \tilde{H}_{\mathrm{XY5Q}}(t) =\sum_{\substack{j=1 \\ j \neq 2}}^{5}  J( e^{i \Delta_{j} t }\hat{\sigma}_{j}^{+}\hat{\sigma}_{2}^{-} +  e^{-i \Delta_{j} t }\hat{\sigma}_{j}^{-}\hat{\sigma}_{2}^{+}),
\end{align}
where $\Delta_{j} = \omega_{j} - \omega_{2}$ denotes the frequency detuning between $Q_{2}$ and its neighboring qubit $Q_{j}$.

Since all parameters for performing FM and DD, including the optimal modulation amplitude $\gamma_{\rm opt}^{N}$, the matched gate time $T_M$, and the length of the DD sequence, are determined by the magnitude of the detuning $|\Delta|$, we consider a symmetric case in which all neighbors of $Q_{2}$ share the same detuning $\Delta$. Under this condition, the pulse designs developed for the two-qubit model become directly applicable to the five-qubit system. This construction can be further generalized to a two-dimensional lattice explicitly labeled by $Q_{i,j}$, where a uniform detuning pattern can be imposed such that $\omega_{i+1,j} - \omega_{i,j}=\omega_{i,j+1} - \omega_{i,j}=\Delta$ for all lattice indices $(i,j)$. Such a frequency arrangement establishes a systematic approach for suppressing XY crosstalk across large-scale qubit arrays.

\subsection{Scalable crosstalk suppression under frequency modulation}
\begin{figure}[htbp]
\begin{center}
\includegraphics[width=1\linewidth]{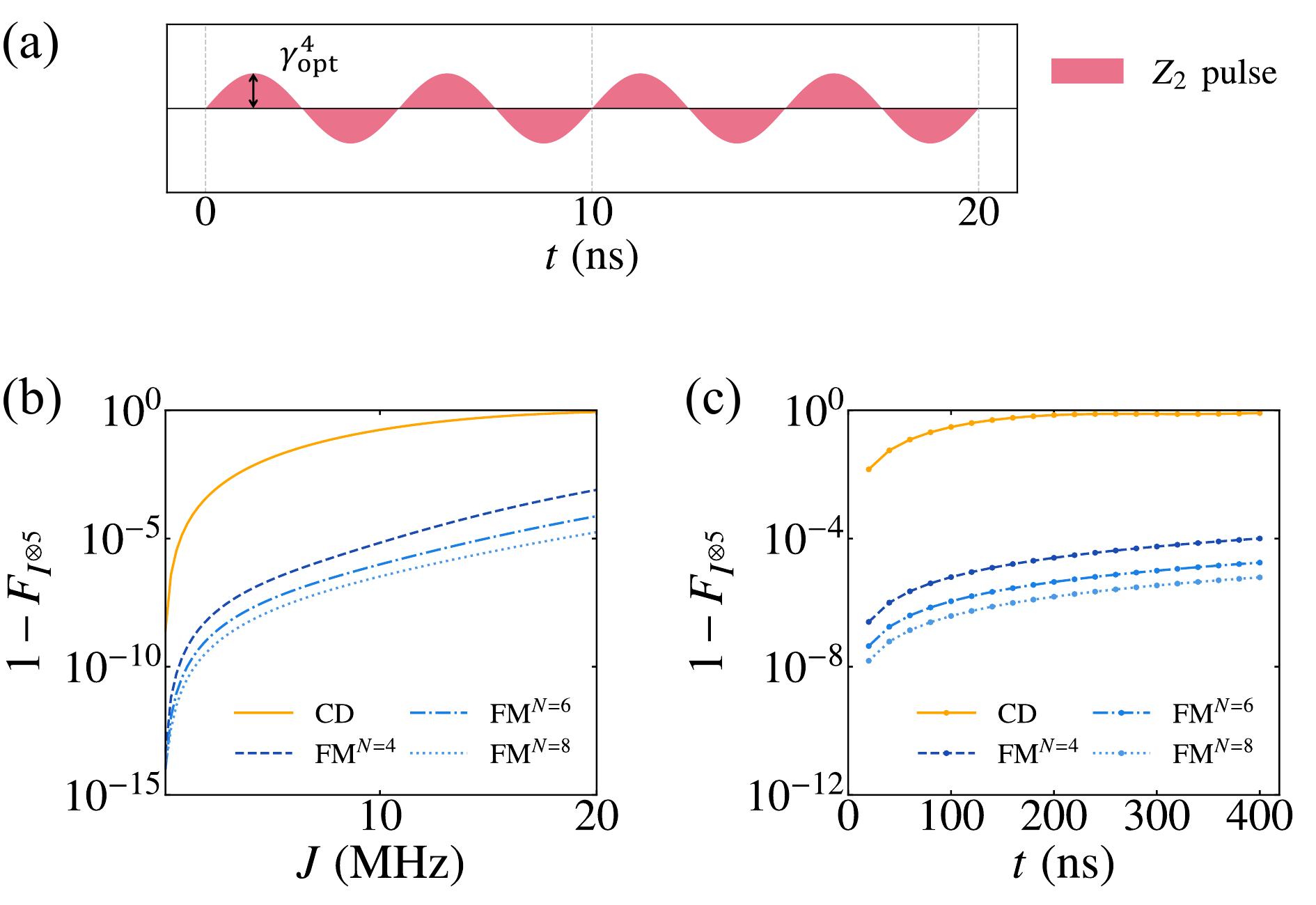}
\caption{Suppression of idle gate infidelity in a five-qubit subsystem under frequency modulation.
(a) Waveform of a sinusoidal $Z_2$ drive with $N=4$ modulation cycles applied to $Q_2$.
(b) Idle gate infidelity of one idle operation as a function of the coupling strength $J$ for different numbers of modulation cycles $N$.
(c) Idle gate infidelity of consecutive idle operations as a function of time for different numbers of modulation cycles $N$, with $J/2\pi = 5$~MHz.}
\label{fig:FM5qidle}
\end{center}
\end{figure}

We evaluate the performance and scalability of frequency modulation in suppressing XY crosstalk during both idle and single-qubit $X_{2}$ operations in the five-qubit configuration. For the idle gate, an example of the applied $Z$-modulation corresponding to the $N=4$ case is shown in Fig.~\ref{fig:FM5qidle}(a), which is identical to the two-qubit case shown in Fig.~\ref{fig:FMidleInfidelity}(a). For $\Delta/2\pi = 50$~MHz and the matched gate time $T_M = 20$~ns, FM reduces the simulated idle gate infidelity by more than four orders of magnitude relative to crosstalk dynamics for $J/2\pi \ge 5$~MHz, as shown in Fig.~\ref{fig:FM5qidle}(b). Although the absolute infidelity is higher than in the two-qubit case due to the additional coupling paths, the suppression achieved by FM remains significant.

We further simulate repeated idle operations under frequency modulation. As shown in Fig.~\ref{fig:FM5qidle}(c), after 20 consecutive idle gates, the FM-protected infidelity remains approximately four orders of magnitude below that under CD. The suppression during repeated operations is comparable to the two-qubit case (Fig.~\ref{fig:FMidleInfidelity}(c)), indicating that FM maintains its effectiveness in larger qubit systems.

\begin{figure}[htbp]
\begin{center}
\includegraphics[width=1\linewidth]{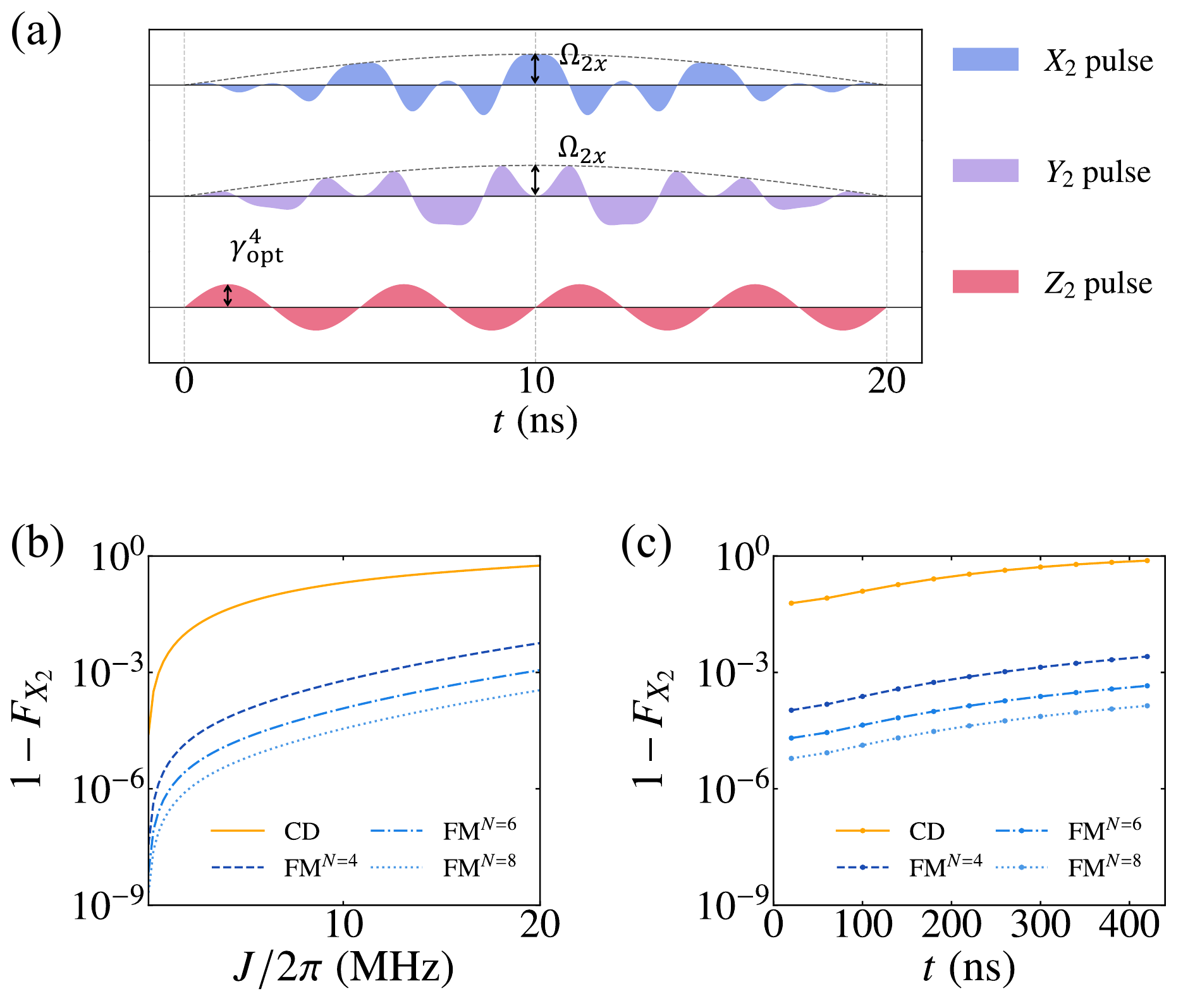} 
\caption{Suppression of $X_2$ gate infidelity in a five-qubit subsystem under frequency modulation. (a) Waveforms of the $X_2$ and $Y_2$ control pulses for generating the $X_2$ gate, and the sinusoidal $Z_2$ drive with $N=4$ modulation cycles applied to $Q_2$. (b) $X_{2}$ gate infidelity of one $X_2$ operation as a function of the coupling strength $J$ for different numbers of modulation cycles $N$.
(c) $X_{2}$ gate infidelity of consecutive $X_2$ operations as a function of time for different numbers of modulation cycles $N$, with $J/2\pi = 5$~MHz.}
    \label{fig:FM5qX1}
\end{center}
\end{figure}

For active gate operation, we consider a single-qubit $X_{2}$ gate applied to $Q_{2}$, the central qubit in the five-qubit layout. For scalable construction, we adopt the single-site scheme in which both the $X_{2}$ drive and the $Z_{2}$ modulation act on the same qubit. As a result, the pulse sequence differs from the two-qubit case, as detailed in Appendix~\ref{app:SingleQFM}. An example of the frequency-modulated $Z_{2}$ pulse together with the $X_{2}$ and $Y_{2}$ control pulses used to generate the $X_{2}$ gate for the $N=4$ case is shown in Fig.~\ref{fig:FM5qX1}(a). As seen in Fig.~\ref{fig:FM5qX1}(b), FM reduces the $X_{2}$-gate infidelity by more than two orders of magnitude for $J/2\pi \ge 5$~MHz. Although the reduction is slightly smaller than in the two-qubit model, the suppression trend remains consistent.

We then investigate the performance of frequency modulation during repeated gate operations by applying an odd number of $X_{2}$ gates, up to 21 consecutive operations with duration $T_{M}$.
As shown in Fig.~\ref{fig:FM5qX1}(c), the FM-protected gate infidelity remains more than two orders of magnitude lower than that under CD, indicating that FM effectively suppresses the accumulation of errors over extended sequences.

In the five-qubit system, the overall suppression achieved by frequency modulation is slightly reduced compared with the two-qubit case, likely due to the increased number of crosstalk coupling paths. Nevertheless, FM continues to suppress both idle and single-qubit $X$ gate infidelities by several orders of magnitude, maintaining the same qualitative behavior observed in the two-qubit analysis. These results confirm that FM remains effective when extended to multi-qubit systems.

\subsection{Scalable crosstalk suppression under dynamical decoupling}
\begin{figure}[htbp]
\begin{center}
\includegraphics[width=1\linewidth]{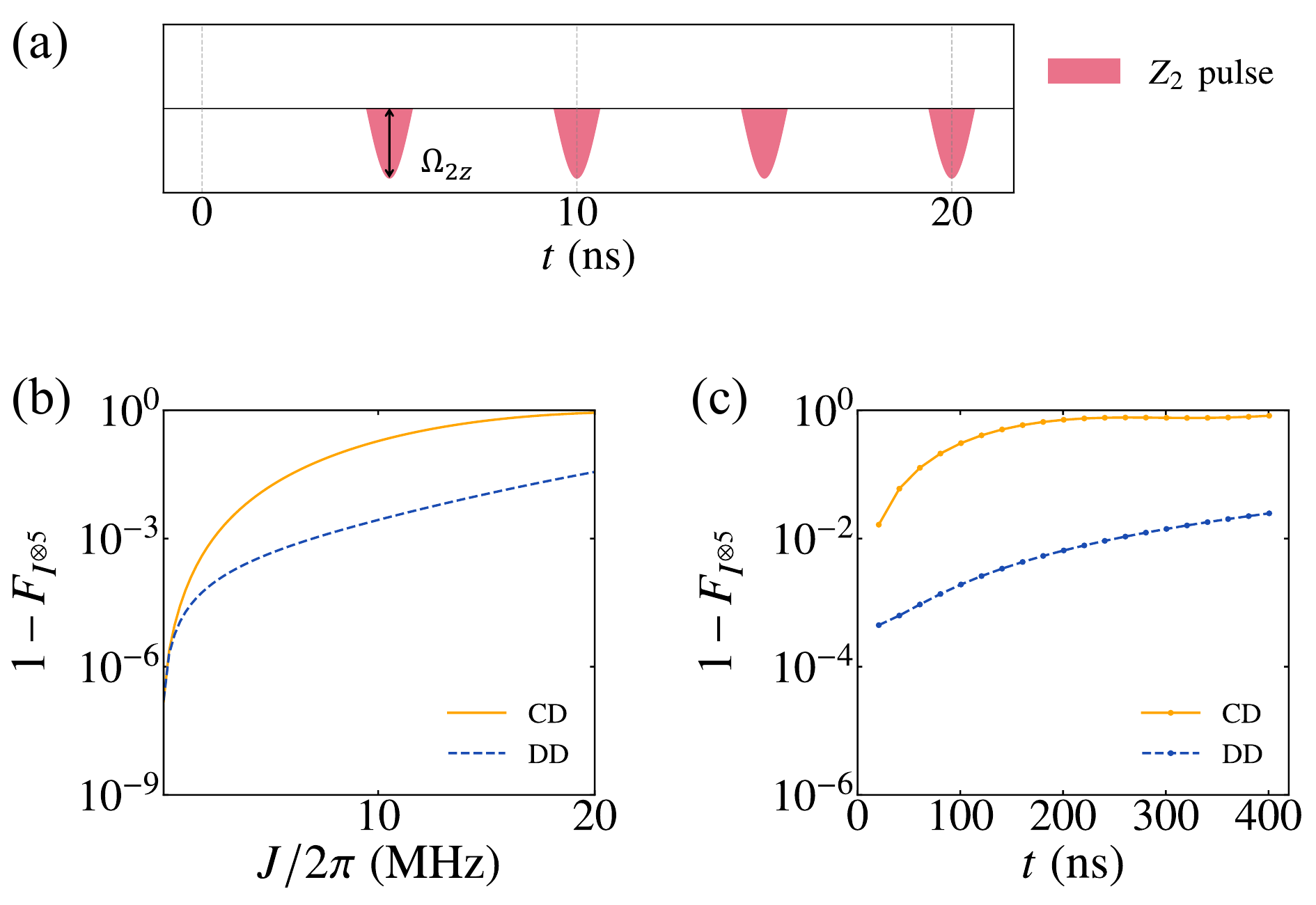}
\caption{Suppression of idle gate infidelity in a five-qubit subsystem under dynamical decoupling.
(a) Waveform of the DD Z-4 sequence containing four $Z_2$ gates applied to $Q_2$.
(b) Idle gate infidelity of one idle operation as a function of the coupling strength $J$ under DD Z-4.
(c) Idle gate infidelity of consecutive idle operations as a function of time under DD Z-4, with $J/2\pi = 5$~MHz.}
\label{fig:DD5qidle}
\end{center}
\end{figure}

Finally, we turn to the performance of dynamical decoupling in suppressing XY crosstalk within the five-qubit configuration. Using the same detuning of $\Delta/2\pi = 50$~MHz as in the previous analysis, we apply the DD Z-4 sequence to $Q_{2}$ during the idle gate, as illustrated in Fig.~\ref{fig:DD5qidle}(a). As shown in Fig.~\ref{fig:DD5qidle}(b), DD Z-4 reduces the idle gate infidelity by more than one order of magnitude at $J/2\pi = 5$~MHz. This reduction is greater than that observed in the two-qubit case (Fig.~\ref{fig:DDidleInfidelity}(b)), indicating enhanced suppression in the multi-qubit configuration. However, the absolute idle gate infidelity remains higher due to the increased number of coupling paths in the five-qubit system. Under repeated idle operations, Fig.~\ref{fig:DD5qidle}(c) shows that DD Z-4 continues to suppress the idle gate error by more than one order of magnitude after 20 consecutive gates.

For single-qubit gate operations, we apply the $X_{2}$ gate to $Q_{2}$ using the pulse sequence shown in Fig.~\ref{fig:DD5qX1}(a), with details provided in Appendix~\ref{app:SingleQDD}. As seen in Fig.~\ref{fig:DD5qX1}(b), the DD Z-4 sequence reduces the $X_{2}$-gate infidelity by approximately $0.37$ orders of magnitude for $J/2\pi \ge 5$~MHz, consistent with the two-qubit results. As in the idle-case analysis, the absolute infidelity is higher in the five-qubit system owing to additional crosstalk paths.

\begin{figure}[htbp]
\begin{center}
\includegraphics[width=1\linewidth]{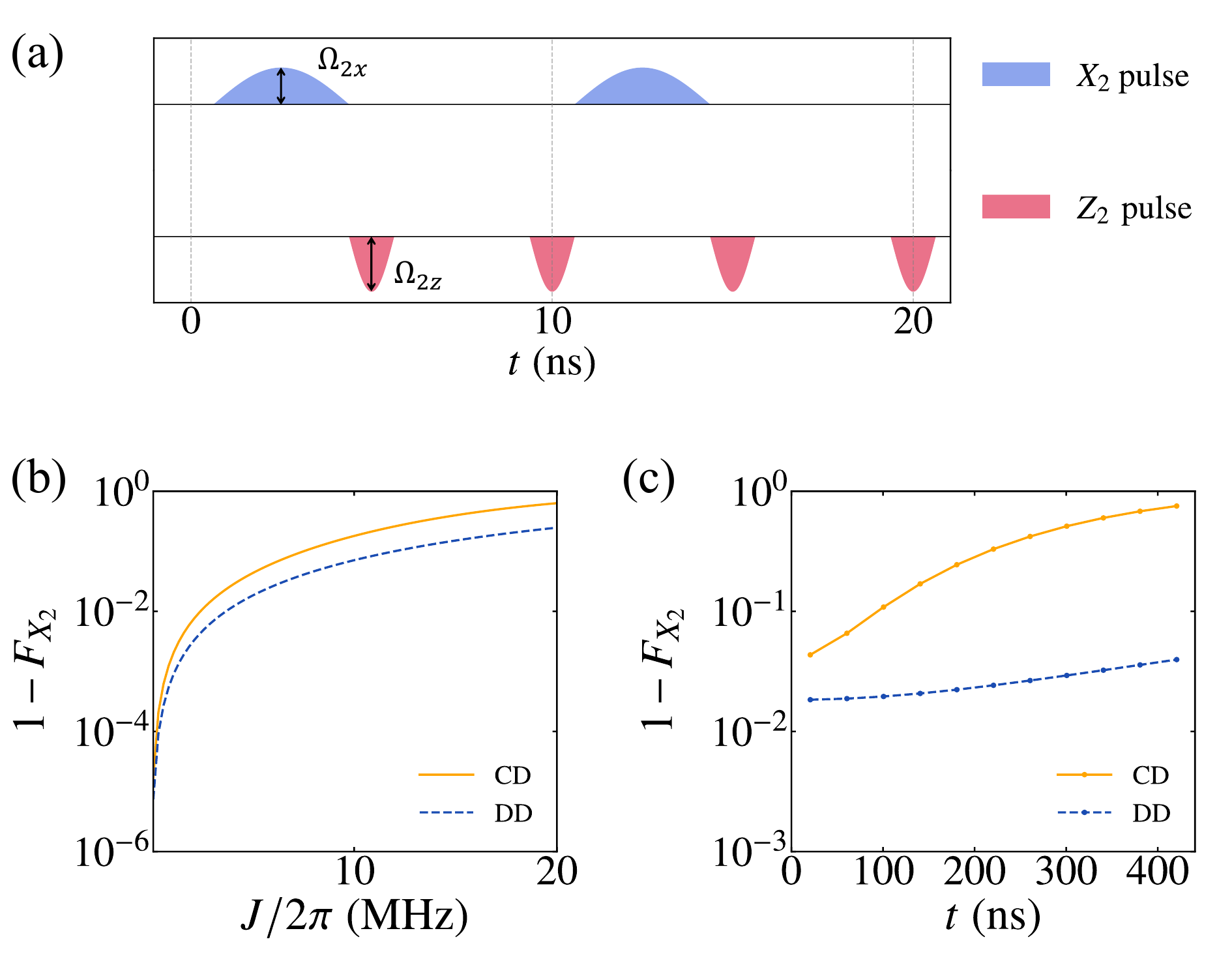}
\caption{Suppression of $X_{2}$ gate infidelity in a five-qubit subsystem under dynamical decoupling.
(a) Waveforms of the $X_2$ control pulses (generated by two $\sqrt{X_2}$ pulses) and the DD Z-4 sequence containing four $Z_2$ gates applied to $Q_2$.
(b) $X_{2}$ gate infidelity of one $X_2$ operation as a function of the coupling strength $J$ under DD Z-4.
(c) $X_{2}$ gate infidelity of consecutive $X_2$ operations as a function of time under DD Z-4, with $J/2\pi = 5$~MHz.}
\label{fig:DD5qX1}
\end{center}
\end{figure}

We further simulate repeated $X_{2}$ operations under DD protection. As shown in Fig.~\ref{fig:DD5qX1}(c), applying up to 21 consecutive $X_{2}$ gates results in more than one order-of-magnitude reduction in infidelity relative to CD. This trend aligns with the two-qubit case and indicates that DD effectively slows the growth of crosstalk-induced errors during extended gate sequences.

Overall, dynamical decoupling remains effective in suppressing XY crosstalk in the five-qubit architecture. While the increased crosstalk connectivity leads to higher absolute infidelities compared with the two-qubit model, DD consistently reduces both idle and single-qubit gate errors and provides increasing relative suppression over extended operations. These results confirm that DD, like FM, can be reliably extended to multi-qubit systems.

\section{Discussion}
In this work, we develop pulse-level control strategies to mitigate residual XY crosstalk during single-qubit operations in superconducting qubit architectures. Through phase averaging at matched gate times and extending this principle through frequency modulation and dynamical decoupling, we show that crosstalk-induced errors can be suppressed for both idle periods and active gates. This framework improves the fidelity of standard single-qubit gates and naturally extends to suppress crosstalk from multiple nearby qubits, providing a scalable solution for densely connected architectures without additional hardware overhead.

The two proposed methods offer complementary trade-offs for experimental implementation. Continuous frequency modulation achieves robust error suppression with gate-time flexibility but comes at the expense of potential increased decoherence when operating away from flux sweet spots. In contrast, discrete dynamical decoupling preserves standard gate implementations but faces constraints in fitting sequences within short gate durations. Both methods also require consideration of the heating induced by the flux control, making future experimental validation essential to assess realistic performance.

These results open several avenues for future extension. Incorporating higher qubit levels would broaden the applicability of the proposed control strategies under realistic operating conditions, while architectures with tunable couplers or indirect interactions offer a natural setting for further exploration. Extending crosstalk-suppression strategies beyond single-qubit operations to two-qubit gates remains an important next step. These directions outline a viable pathway toward pulse-level crosstalk mitigation compatible with realistic and scalable superconducting quantum processors.

\begin{acknowledgments}
We thank Chung-Ting Ke, Watson Kuo, and Li-Chieh Hsiao for helpful discussions. C.-H.~Wang acknowledges funding support from the National Science and Technology Council, Taiwan, under grant numbers 111-2112-M-002-049-MY3, 114-2119-M-007-013-, 114-2124-M-002-003-, and 114-2112-M-002-021-MY3, and from the Office of Research and Development, National Taiwan University, under grant numbers~\mbox{114L895001} and ~\mbox{115L893701}. C.-H. Wang is also grateful for the NTU Eminence Scholar Fellowship funded by the Fubon Foundation and the Taiwan Semiconductor Manufacturing Company, as well as the support from the Physics Division, National Center for Theoretical Sciences, Taiwan.
\end{acknowledgments}

\appendix
\section{Detailed analysis of frequency modulation\label{App:FMdetails}}
\subsection{Optimization for the idle gate \label{App:FMIdle}}
For a matched gate time $T_M$, the first-order contribution from the XY term vanishes, as in the crosstalk dynamics case [Eq.~\eqref{eqn:HCD1}]. We therefore examine how frequency modulation further suppresses second-order error contributions, and identify the optimal modulation amplitude $\gamma^{N}_{\mathrm{opt}}$ that achieves strong suppression.
The corresponding second-order Magnus term for the idle gate under FM reads
\begin{align}
    &\bar{H}^{\mathrm{FM}(2)}_{\mathrm{idle}}=-\frac{i}{2T} \int _{0}^{T} \mathrm{d}t_{1} \int _{0}^{t_1} \mathrm{d}t_{2} \left[\tilde{H}_{\mathrm{XY}}^{V}(t_{1}),\tilde{H}_{\mathrm{XY}}^{V}(t_{2}) \right] \notag \\ &=  \frac{J^2}{2T} \int _{0}^{T} \mathrm{d}t_{1} \int _{0}^{t_1} \mathrm{d}t_{2}    E^{(2)}_{\mathrm{idle}}(t_1,t_2)(\hat{\sigma}_{1}^{z}\hat{I}_{2}- \hat{I}_{1} \hat{\sigma}^{z}_{2}),
    \label{eqn:HFM2idle}
\end{align}
where $E^{(2)}_{\mathrm{idle}}(t_1,t_2) =\sin\left[\Delta (t_1 - t_2)+2(\alpha(t_1) - \alpha(t_2)) \right] $.

\begin{figure}[htbp]
    \begin{center}
\includegraphics[width=0.7\linewidth]{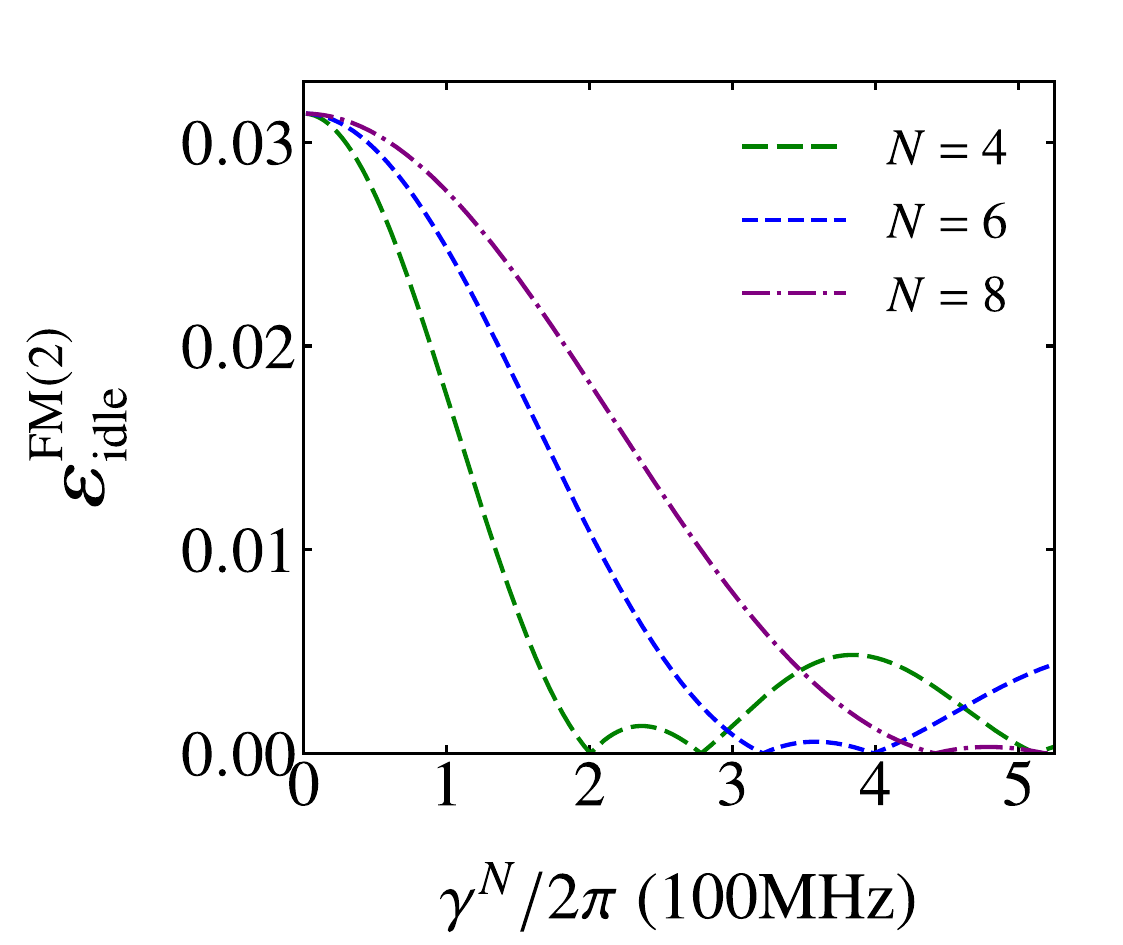}
\caption{Second-order idle gate crosstalk error under frequency modulation, $\varepsilon^{\mathrm{FM}(2)}_{\mathrm{idle}}$, as a function of the modulation amplitude $\gamma^{N}$ for different numbers of modulation cycles $N$. The chosen parameters are $T_M=20$~ns, $\Delta/2\pi=50$~MHz, and $J/2\pi=5$~MHz.}
    \label{fig:eFM2idle}
    \end{center}
\end{figure}

The second-order crosstalk error for the idle gate under frequency modulation is quantified as the sum of the magnitudes of the coefficients preceding the corresponding second-order operators,
\begin{align}
     &\varepsilon^{\rm{FM}(2)}_{\rm{idle}}=\frac{J^2}{T} \abs{\int_{0}^{T} \mathrm{d} t_{1} \int _{0}^{t_1} \mathrm{d} t_{2} \, E^{(2)}_{\mathrm{idle}}(t_1,t_2)}.
     \label{eqn:eFM2idle}
\end{align}
For each modulation cycle number $N$, we numerically evaluate the second-order crosstalk error under different values of $\gamma^{N}$ and identify the optimal modulation amplitude $\gamma^{N}_{\mathrm{opt}}$ for a matched gate time $T_M = 20$~ns, as shown in Fig.~\ref{fig:eFM2idle}. We select the value of $\gamma^{N}$ corresponding to the first local minimum of $\varepsilon^{\mathrm{FM}(2)}_{\mathrm{idle}}$ as $\gamma^{N}_{\mathrm{opt}}$, since lower-frequency modulation is more practical for hardware implementation.

We find that the numerically evaluated second-order crosstalk error for the idle gate $\varepsilon^{\mathrm{FM}(2)}_{\mathrm{idle}}$ exhibits corner-like minima. Therefore, in the simulations presented in Sec.~\ref{subsec:FMIdle} of the main text, we do not directly use the numerically identified minimum at $\gamma^{N}_{\mathrm{opt}}$. Instead, we define the idle gate infidelity using a numerically stable estimate evaluated by averaging the values at $\gamma^{N}_{\mathrm{opt}} \pm \Delta\gamma$,
\begin{equation}
F^{\mathrm{FM}}_{I_{1}I_{2}}
\equiv
\frac{1}{2}\left[
F^{\mathrm{FM}}_{I_{1}I_{2}}(\gamma^{N}_{\mathrm{opt}} + \Delta\gamma)
+
F^{\mathrm{FM}}_{I_{1}I_{2}}(\gamma^{N}_{\mathrm{opt}} - \Delta\gamma)
\right],
\end{equation}
where $\Delta\gamma$ denotes the sampling step used in the numerical scan of $\gamma^{N}$. For the parameters used in Fig.~\ref{fig:eFM2idle}, the resulting optimal modulation amplitudes are $\gamma^{4}_{\mathrm{opt}}/2\pi = 201~ \mathrm{MHz}$ , $\gamma^{6}_{\mathrm{opt}}/2\pi = 321~\mathrm{MHz}$, and $\gamma^{8}_{\mathrm{opt}}/2\pi = 442~ \mathrm{MHz}$, with $\Delta \gamma/2\pi=1.59~\mathrm{MHz}$.

\subsection{Optimization for the $X_1$ gate \label{App:FMX1}}
In contrast to the idle-gate case, the second-order contribution for the $X_1$ gate includes additional commutators between the drive term and the XY crosstalk. The Hamiltonian for generating a single-qubit $X_1$ gate in the presence of crosstalk, expressed in the frequency-modulated frame, is given by
\begin{align}
    \tilde{H}^{V}(t) =& \tilde{H}_{\mathrm{XY}}^{V}(t) + \tilde{H}^{V}_{\mathrm{drive}}(t) \notag \\ = &J \left( e^{i \left[\Delta t +2\alpha(t)\right]}\hat{\sigma}_{1}^{+}\hat{\sigma}_{2}^{-} +  e^{-i \left[\Delta t +2\alpha(t)\right]}\hat{\sigma}_{1}^{-}\hat{\sigma}_{2}^{+} \right) \notag \\ &+ \Omega_{1}(t) \hat{\sigma}_{1}^{x},
\end{align}
where the driving pulse is chosen as $ \Omega_{1}(t) =\Omega_{1x}\sin(\frac{\pi}{T}t)$ with $\Omega_{1x}=\frac{\pi^2}{4T}$, satisfying $\int_{0}^{T} \Omega_{1}(t)\mathrm{d}t = \frac{\pi}{2} $.

The second-order Magnus term for the $X_1$ gate operation under frequency modulation is given by
\begin{align}
    \bar{H}^{\mathrm{FM}(2)}_{X_{1}} = - \frac{i}{2T} \int^{T}_{0}\mathrm{d}t_{1} \int^{t_{1}}_{0}\mathrm{d}t_{2} [\tilde{H}^{V}(t_1), \tilde{H}^{V}(t_2)].
\end{align}
The commutator can be decomposed into four components,
\begin{align} &[\tilde{H}^{V}(t_1), \tilde{H}^{V}(t_2)]\notag  \\&= \left[\tilde{H}_{\mathrm{XY}}^{V}(t_{1}), \tilde{H}_{\mathrm{XY}}^{V}(t_{2}) \right] + \left[\tilde{H}^{V}_{\mathrm{drive}}(t_1),\tilde{H}^{V}_{\mathrm{drive}}(t_2) \right] \notag \\
   & + \left[\tilde{H}^{V}_{\mathrm{drive}}(t_1), \tilde{H}_{\mathrm{XY}}^{V}(t_{2}) \right]+ \left[\tilde{H}_{\mathrm{XY}}^{V}(t_{1}), \tilde{H}^{V}_{\mathrm{drive}}(t_2) \right].
\end{align}

The first component arises solely from the XY crosstalk and is identical to that in Eq.~\eqref{eqn:HFM2idle}. The corresponding error has already been evaluated in the idle-gate case [Eq.~\eqref{eqn:eFM2idle}]. The second component corresponds to the ideal drive term and does not contribute to crosstalk-induced errors.
We thus focus on the remaining two components, which involve the commutators between the drive and the XY crosstalk. The sum of these two components can be expressed as
\begin{align}
    &\left(-\Omega_{1}(t_{1})Je^{i\left[\Delta t_{2} + 2\alpha(t_2)\right]}+\Omega_{1}(t_{2})Je^{i\left[\Delta t_{1} + 2\alpha(t_1)\right]}\right)  \hat{\sigma}^{z}_{1}\hat{\sigma}^{-}_{2} \notag \\ &+\left(\Omega_{1}(t_{1})Je^{-i\left[\Delta t_{2} + 2\alpha(t_2)\right]}- \Omega_{1}(t_{2})Je^{-i\left[\Delta t_{1} + 2\alpha(t_1)\right]} \right)\hat{\sigma}^{z}_{1}\hat{\sigma}^{+}_{2}.
    \label{eqn:FM(2)X1commutator}
\end{align}

Since the two operators in Eq.~\eqref{eqn:FM(2)X1commutator} have the same coefficient magnitudes, the total second-order error is quantified as
\begin{align}
     &\varepsilon^{\mathrm{FM}(2)}_{X_1} =  \left| \frac{iJ}{T} \int _{0}^{T} \mathrm{d}t_{1} \int _{0}^{t_1} \mathrm{d}t_{2}\,E^{\mathrm{FM}(2)}_{X_{1}}(t_{1}, t_{2})\right| +\varepsilon^{\mathrm{FM}(2)}_{\mathrm{idle}}, \notag \\ &E^{\rm FM(2)}_{X_{1}}(t_{1}, t_{2}) =\Omega_{1}(t_{1})e^{i\left[\Delta t_2 + 2\alpha(t_2)\right]}-\Omega_{1}(t_2)e^{i\left[\Delta t_{1} + 2\alpha(t_1)\right]}.
     \label{eqn:eFM2X1}
\end{align}

\begin{figure}[htbp]
    \begin{center}    \includegraphics[width=0.7\linewidth]{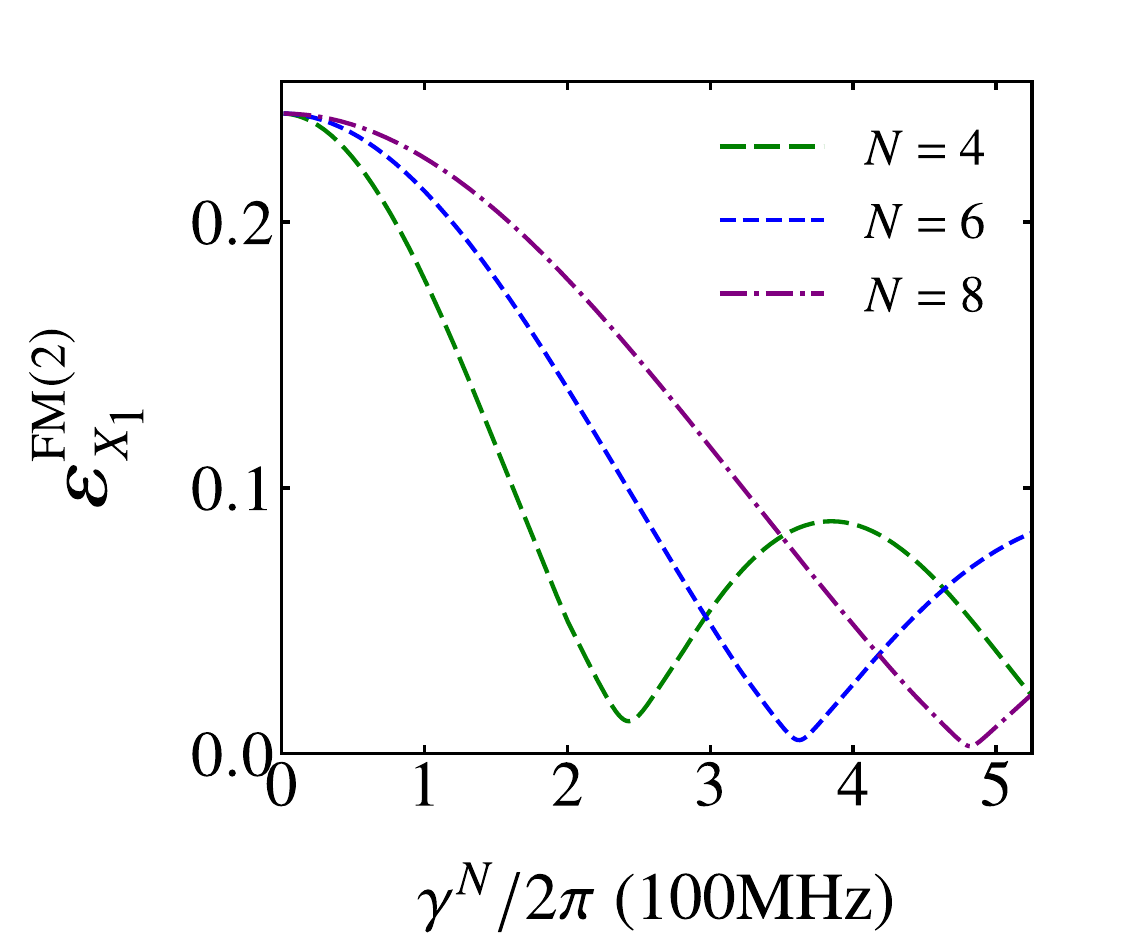}
    \caption{Second-order $X_1$ gate crosstalk error under frequency modulation, $\varepsilon^{\mathrm{FM}(2)}_{X_1}$, as a function of the modulation amplitude $\gamma^{N}$ for different numbers of modulation cycles $N$. The chosen parameters are $T_M=20$~ns, $\Delta/2\pi=50$~MHz, and $J/2\pi=5$~MHz.}
    \label{fig:eFM2X1}
    \end{center}
\end{figure}

For a given matched gate time $T_M=20$~ns, we numerically evaluate the magnitude of $\varepsilon^{\mathrm{FM}(2)}_{X_1}$ for each modulation cycle number $N$, as shown in Fig.~\ref{fig:eFM2X1}. We identify the optimal modulation amplitude $\gamma^{N}_{\mathrm{opt}}$ by selecting the value of $\gamma^{N}$ corresponding to the first local minimum of $\varepsilon^{\mathrm{FM}(2)}_{X_1}$ as $\gamma^{N}_{\mathrm{opt}}$. The resulting optimal modulation amplitudes are $\gamma^{4}_{\mathrm{opt}}/2\pi =244~\mathrm{MHz}$, $\gamma^{6}_{\mathrm{opt}}/2\pi=363~\mathrm{MHz}$, and $\gamma^{8}_{\mathrm{opt}}/2\pi=482~\mathrm{MHz}$, which is used in the simulations presented in Sec.~\ref{subsec:FMX1}.

\subsection{Single-site scheme of frequency modulation \label{app:SingleQFM}}
In practical implementations, it is often desirable to apply both the control drive and the frequency modulation to the same qubit. We therefore consider the case where both the drive and $Z$-modulation are applied to the same qubit $Q_2$ to achieve single-qubit operations and crosstalk suppression. In the presence of $Z$-modulation, the action of the drive Hamiltonian is modified. To realize the desired single-qubit $X_2$ gate in the static rotating (operation) frame, the drive term must therefore be adjusted correspondingly. Specifically, the drive Hamiltonian in the frequency-modulated frame is related to that in the operation frame by
\begin{align}
    \tilde{H}^{V}_{\mathrm{drive}}(t)=e^{i[\alpha(t)]\hat{\sigma}^{z}_{2}}\tilde{H}_{\mathrm{drive}}(t)e^{-i[\alpha(t)]\hat{\sigma}^{z}_{2}},
    \label{eqn:drivetransform}
\end{align}
where $\alpha(t) = \frac{\gamma T}{\pi N} \sin^{2}{(\frac{\pi N}{T}t)}$.

To realize a single-qubit $X_2$ gate, the drive Hamiltonian in the frequency-modulated frame is chosen as $\tilde{H}^{V}_{\mathrm{drive}}(t) = \Omega_{2}(t)\hat{\sigma}_{2}^{x}$ with a smooth envelope $\Omega_{2}(t) =\Omega_{2x}\sin(\tfrac{\pi}{T}t)$, where $\Omega_{2x}=\frac{\pi^2}{4T}$ is set such that $\int_{0}^{T}\Omega_{2}(t) \mathrm{d}t=\frac{\pi}{2}$. The corresponding drive Hamiltonian in the operation frame is then given by
\begin{align}
    \tilde{H}_{\mathrm{drive}}(t) =\Omega_{2}(t)\left[\cos{\left(2\alpha(t)\right)} \hat{\sigma}_{2}^{x} + \sin{\left(2\alpha(t)\right)} \hat{\sigma}_{2}^{y} \right],
\end{align}
obtained via the inverse transformation of Eq.~\eqref{eqn:drivetransform}.

Following the derivation for the $X_1$ gate in Appendix~\ref{App:FMX1}, we obtain the same second-order crosstalk error expression for the single-site $X_2$ gate. In particular, since the drive is chosen to have the same form in the frequency-modulated frame, the resulting error $\varepsilon^{\mathrm{FM}(2)}_{X_2}$ is identical to $\varepsilon^{\mathrm{FM}(2)}_{X_1}$ given in Eq.~\eqref{eqn:eFM2X1}. Consequently, the optimal modulation amplitudes $\gamma^{N}_{\mathrm{opt}}$ are the same as those obtained in Appendix~\ref{App:FMX1}.

Figure~\ref{fig:FMX2Infidelity}(a) shows an example of the control waveforms for the single-site frequency modulation protocol with $N=4$ at the matched gate time $T_M=20$~ns. In contrast to the two-site scheme, frequency modulation on the same qubit modifies the effective drive, such that the target $X_2$ gate is realized using combined $X_2$ and $Y_2$ drives, while the sinusoidal $Z_2$ modulation suppresses the residual XY interaction.

\begin{figure}[htbp]
\begin{center}
\includegraphics[width=1\linewidth]{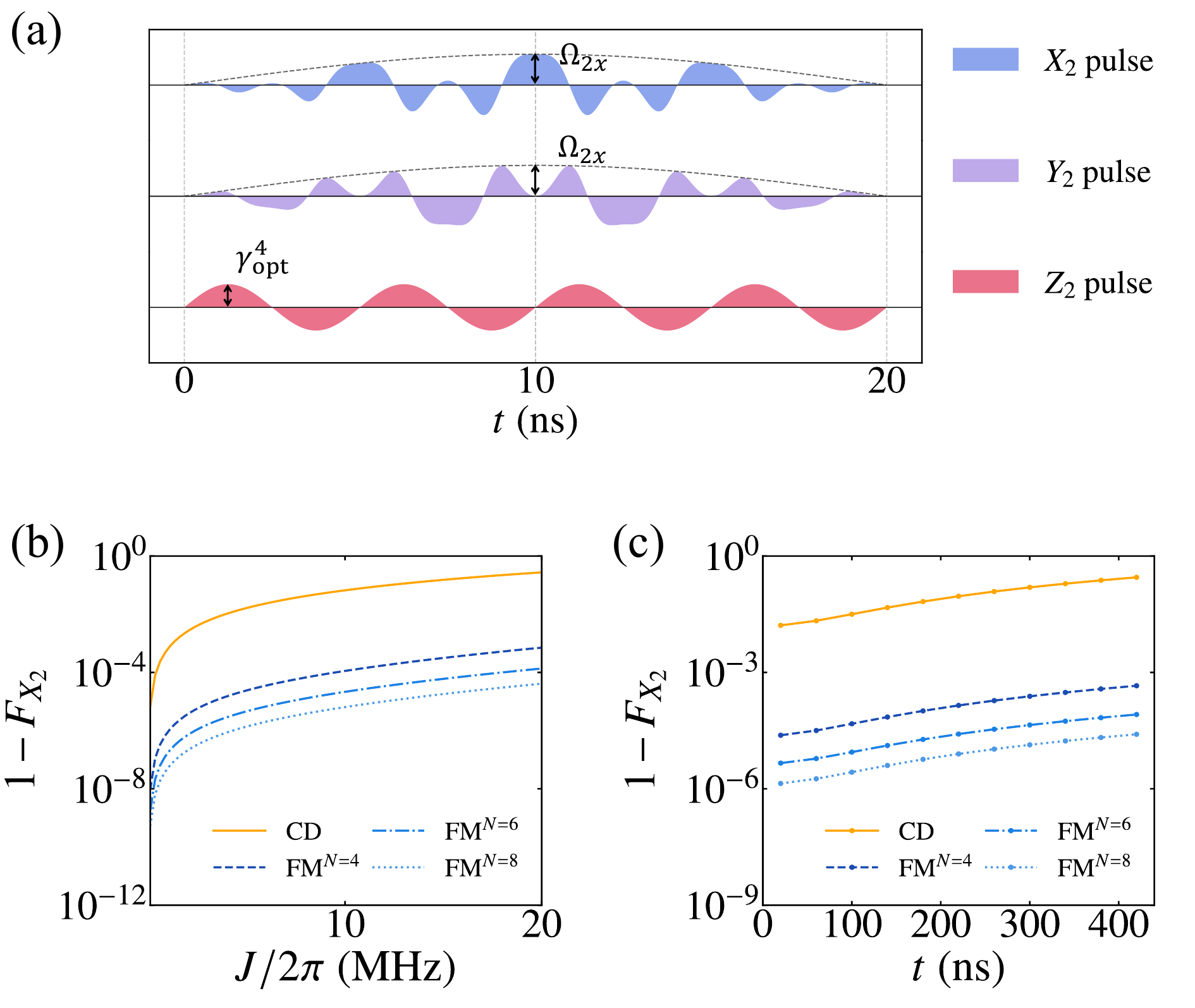} 
\caption{Suppression of $X_{2}$ gate infidelity under single-site frequency modulation. (a) Waveforms of the $X_2$ and $Y_2$ control pulses for generating the $X_2$ gate, and the sinusoidal $Z_2$ drive with $N=4$ modulation cycles applied to $Q_2$. (b) $X_{2}$ gate infidelity of one $X_2$ operation as a function of the coupling strength $J$ for different numbers of modulation cycles $N$. (c) $X_{2}$ gate infidelity of consecutive $X_2$ operations as a function of time for different numbers of modulation cycles $N$, with $J/2\pi=5$~MHz.}
\label{fig:FMX2Infidelity}
\end{center}
\end{figure}

As shown in Fig.~\ref{fig:FMX2Infidelity}(b), single-site frequency modulation reduces the infidelity of $X_2$ gate by more than two orders of magnitude compared with the crosstalk dynamics case.
We further simulate sequences of consecutive $X_2$ gates to assess the long-term performance.
As shown in Fig.~\ref{fig:FMX2Infidelity}(c), even after 21 consecutive operations, the infidelity remains more than two orders of magnitude lower than without FM.
These results are consistent with those obtained for the $X_1$ gate (Fig.~\ref{fig:FMX1Infidelity}), confirming that both the control drive and the $Z$ modulation can be applied to the same qubit to achieve robust suppression of XY crosstalk, provided that the drive waveform is appropriately modified.

\subsection{Parallel $X_1X_2$ operations under frequency modulation \label{app:X1X2FM}}
In the main text (Sec.~\ref{subsec:FMX1}), the $Z_2$ modulation is shown to suppress crosstalk errors for $X_1$ gate applied to $Q_1$. Appendix~\ref{app:SingleQFM} further shows that the same $Z_2$ modulation can also protect an $X_2$ gate on $Q_2$, provided that the drive waveform on $Q_2$ is correspondingly modified. We now consider parallel single-qubit operations, where $X_1$ and $X_2$ gates are applied simultaneously to $Q_1$ and $Q_2$, respectively, under the same $Z_2$ modulation.

We consider the same forms of the $X_1$, $X_2$, and $Y_2$ driving pulses as those used in Sec.~\ref{subsec:FMX1} and Appendix~\ref{app:SingleQFM}. Under this condition, the second-order crosstalk error for the parallel $X_1X_2$ operation is given by the sum of the corresponding single-gate errors, with the shared idle error contribution subtracted, $\varepsilon^{\mathrm{FM}(2)}_{X_1 X_2} =\varepsilon^{\mathrm{FM}(2)}_{X_1}+\varepsilon^{\mathrm{FM}(2)}_{X_2} -\varepsilon^{\mathrm{FM}(2)}_{\mathrm{idle}} $, and the optimal modulation amplitudes $\gamma^{N}_{\mathrm{opt}}$ remain identical to those determined previously for the single $X$-gate case.

\begin{figure}[htbp]
    \begin{center}
    \includegraphics[width=1\linewidth]{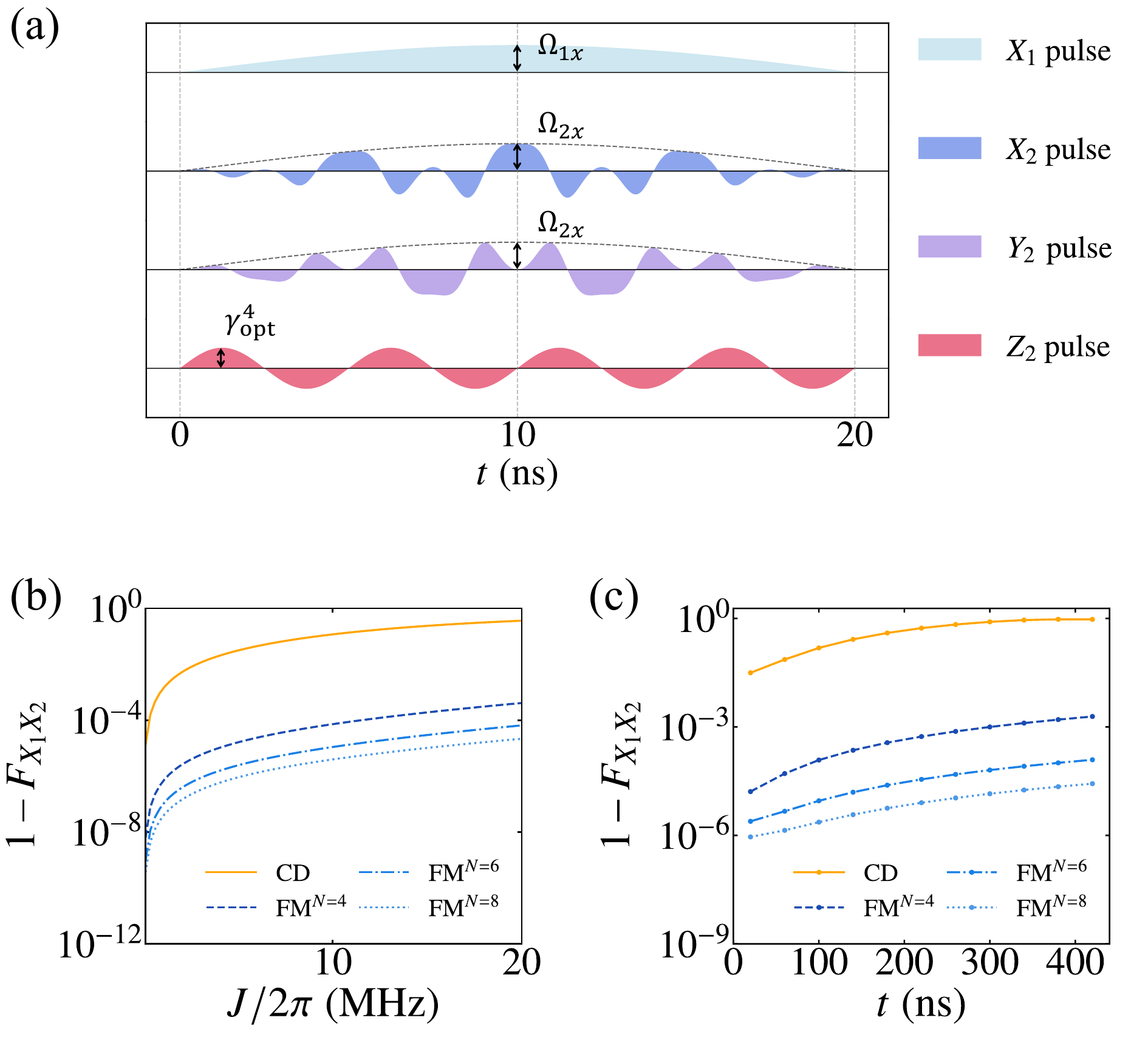} \caption{Suppression of parallel $X_{1}X_{2}$ gate infidelity under frequency modulation. (a) Waveforms of the $X_1$ control pulses applied to $Q_1$, the $X_2$ and $Y_2$ control pulses for generating the $X_2$ gate on $Q_2$, and the sinusoidal $Z_2$ drive with $N=4$ modulation cycles applied to $Q_2$. (b) Parallel $X_{1}X_{2}$ gate infidelity as a function of the coupling strength $J$ for different numbers of modulation cycles $N$. (c) Parallel $X_{1}X_{2}$ gate infidelity of consecutive operations as a function of time for different numbers of modulation cycles $N$, with $J/2\pi=5$~MHz.}
    \label{fig:FMX1X2infidelity}
    \end{center}
\end{figure}

Figure~\ref{fig:FMX1X2infidelity}(a) shows a representative example of the control waveforms for the parallel $X_1X_2$ operations with $N=4$. The $X_1$ drive pulse is identical to that used in Sec.~\ref{subsec:FMX1}, while the $X_2$ and $Y_2$ drive components follow the single-site scheme described in Appendix~\ref{app:SingleQFM}. These pulses are applied simultaneously under the same $Z_2$ modulation.

As shown in Fig.~\ref{fig:FMX1X2infidelity}(b), for the matched gate time $T_M=20$~ns, frequency modulation reduces the infidelity of parallel $X_1X_2$ operations by more than three orders of magnitude compared with crosstalk dynamics. The suppression remains robust under repeated operations: Fig.~\ref{fig:FMX1X2infidelity}(c) shows that even after 21 consecutive parallel gates, the infidelity remains more than two orders of magnitude lower than in the absence of frequency modulation.

\subsection{Frequency modulation with non-matched gate time}
We have shown that frequency modulation can suppress XY crosstalk at the matched gate time $T_M$, where the first-order error vanishes and the remaining second-order contribution is further suppressed. More generally, for a given non-matched gate time $T$, an optimal modulation amplitude $\gamma$ can still be determined by minimizing the first-order crosstalk error,
\begin{align}
    \varepsilon^{\mathrm{FM}(1)}\equiv \left|\frac{J}{T} \int_{0}^{T}  e^{i\left[\Delta t + 2\alpha(t)  \right]}\mathrm{d}t \right| +\left|\frac{J}{T} \int_{0}^{T}  e^{-i \left[ \Delta t + 2\alpha(t)  \right]}\mathrm{d}t \right|.
    \label{eqn:eFM1APP}
\end{align}

\begin{figure}[!htbp]
    \centering
    \includegraphics[width=0.7\linewidth]{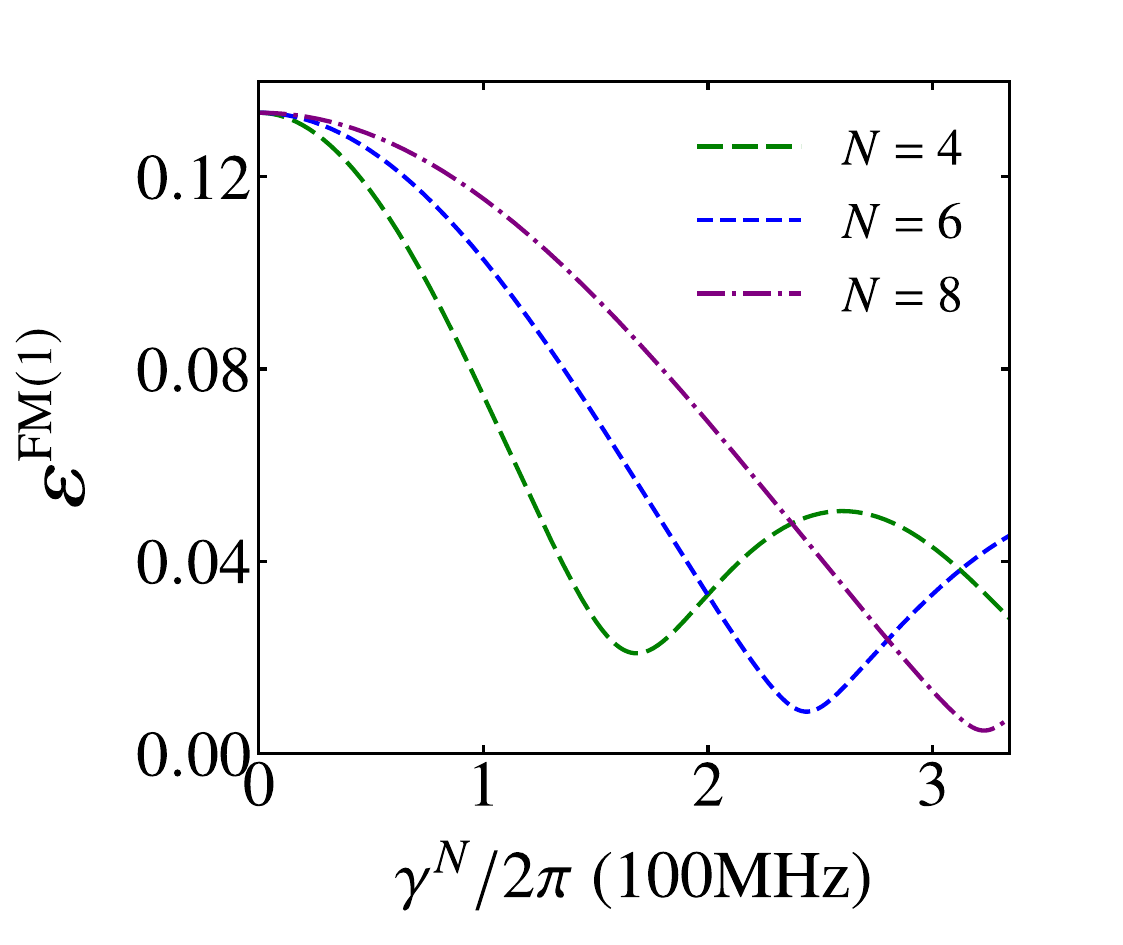}
    \caption{First-order crosstalk error under frequency modulation, 
$\varepsilon^{\mathrm{FM}(1)}$, as a function of the modulation amplitude $\gamma^{N}$ 
for different numbers of modulation cycles $N$ at a non-matched gate time. 
The chosen parameters are $T_U = 3T_{\Delta} = 30$~ns, $\Delta/2\pi = 50$~MHz, and $J/2\pi = 5$~MHz.}
    \label{fig:eFM1}
\end{figure}

We consider the same detuning $\Delta/2\pi=50$~MHz as in the previous analysis and choose a non-matched gate time $T_U=3T_{\Delta}=30$~ns. As shown in Fig.~\ref{fig:eFM1}, $\varepsilon^{\mathrm{FM}(1)}$ exhibits local minima for different modulation cycle numbers $N$. For each $N$, we select the first local minimum to define the optimal modulation amplitudes $\gamma^{4}_{\mathrm{opt}}/2\pi = 168.2~\mathrm{MHz}$, $\gamma^{6}_{\mathrm{opt}}/2\pi=243.8~\mathrm{MHz}$, and $\gamma^{8}_{\mathrm{opt}}/2\pi=322.9~\mathrm{MHz}$.

\begin{figure}[htbp]
\begin{center}
\includegraphics[width=1\linewidth]{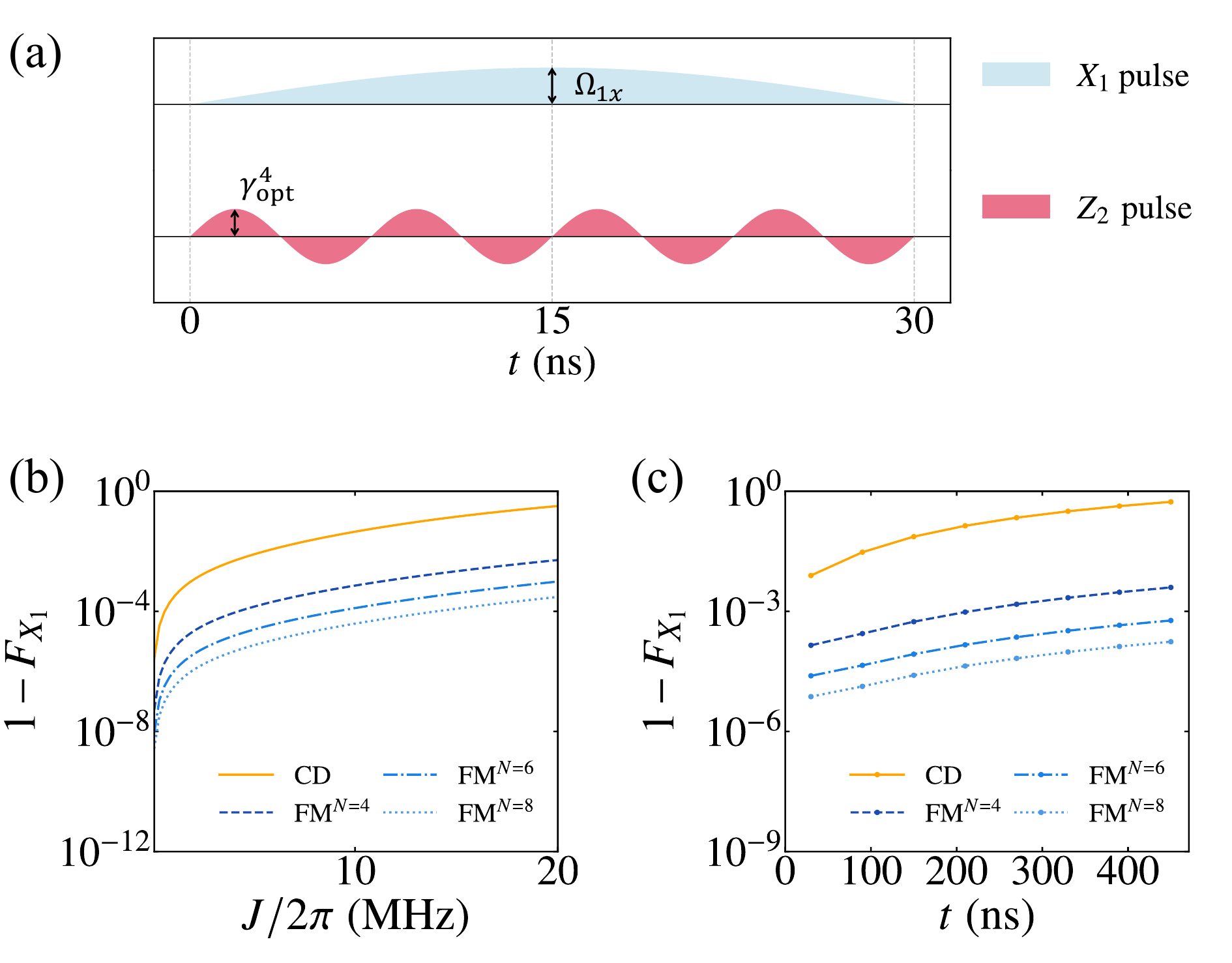} 
\caption{Suppression of $X_{1}$ gate infidelity under frequency modulation at a non-matched gate time. (a) Waveforms of the $X_1$ control pulses applied to $Q_1$ and the sinusoidal $Z_2$ drive with $N=4$ modulation cycles applied to $Q_2$.
(b) $X_{1}$ gate infidelity of one $X_1$ operation as a function of the coupling strength $J$ for different numbers of modulation cycles $N$.
(c) $X_{1}$ gate infidelity of consecutive $X_1$ operations as a function of time for different numbers of modulation cycles $N$. The chosen parameters are $T_U = 3T_{\Delta} = 30$~ns, $\Delta/2\pi = 50$~MHz, and $J/2\pi = 5$~MHz.}
    \label{fig:FMX1_30Infidelity}
\end{center}
\end{figure}

Using the optimized modulation amplitudes $\gamma^{N}_{\mathrm{opt}}$, we
evaluate the performance of frequency modulation for a single-qubit $X_1$ gate at a non-matched gate time in the presence of XY crosstalk. We consider the same form of drive used to generate the $X_1$ gate in the matched gate time case discussed in Sec.~\ref{subsec:FMX1}.
The gate is generated by the driving Hamiltonian
$\tilde{H}^{V}_{\mathrm{drive}}(t)=\Omega_{1}(t)\hat{\sigma}_{1}^{x}$ with a
sinusoidal envelope
$\Omega_{1}(t)=\Omega_{1x}\sin\!\left(\frac{\pi}{T}t\right)$, where
$\Omega_{1x}=\frac{\pi^{2}}{4T}$.
An example of the corresponding control waveform for $N=4$ is shown in
Fig.~\ref{fig:FMX1_30Infidelity}(a).

The simulated gate infidelities as functions of the coupling strength are
summarized in Fig.~\ref{fig:FMX1_30Infidelity}(b).
At the non-matched gate time $T_U=30$~ns, frequency modulation reduces the
$X_1$ gate infidelity by about $1.7$ orders of magnitude compared with crosstalk
dynamics for $J/2\pi\ge 5$~MHz, with larger modulation cycle numbers $N$ leading
to further suppression.
For repeated operations, Fig.~\ref{fig:FMX1_30Infidelity}(c) shows that even after
15 consecutive $X_1$ gates, frequency modulation continues to suppress the
accumulated infidelity by approximately $2.1$ orders of magnitude relative to
CD.

\section{Detailed analysis of dynamical decoupling \label{App:DDdetails}}
\subsection{First-order XY crosstalk suppression condition \label{App:DD1sterror}}
We derive the condition for suppressing the first-order XY
crosstalk under a dynamical decoupling (DD) sequence. 
The first-order XY term
$\bar{H}^{\mathrm{DD(1)}}_{\mathrm{XY}}$, given in
Eq.~\eqref{eqn:general1stDD}, is obtained from the Magnus expansion of the time evolution operator $\hat{U}^{\mathrm{DD}}_{\mathrm{gate}}(T,0)$ over the duration $T$ as
\begin{align}
    \bar{H}^{\mathrm{DD}(1)}_{\mathrm{XY}}
    &= \frac{1}{T}\sum_{s =1}^{S}\int_{(s-1)\tau}^{s\tau}(-1)^{s-1}\tilde{H}_{\mathrm{XY}}(t)\,\mathrm{d}t \notag \\
    &= \frac{1}{T}\sum_{s =1}^{S}\int_{(s-1)\tau}^{s\tau}(-1)^{s-1}J( e^{i \Delta t }\hat{\sigma}_{1}^{+}\hat{\sigma}_{2}^{-} + \mathrm{H.c.})\, \mathrm{d}t \notag \\
    &= \frac{J}{T}\sum_{s =1}^{S} (-1)^{s-1}\left(\frac{1}{i\Delta} e^{i \Delta t }\hat{\sigma}_{1}^{+}\hat{\sigma}_{2}^{-} + \mathrm{ H.c.}  \right) \Bigg|^{s\tau}_{(s-1)\tau}.
    \label{eqn:generalDD1}
\end{align}

The first-order crosstalk error is defined as the sum of the magnitudes of the coefficients preceding the first-order operators,
\begin{align}
    \varepsilon^{\mathrm{DD}(1)}\equiv 2\left|\frac{J}{T}\sum_{s=1}^{S}\frac{1}{i\Delta}(-1)^{s-1}\left[ e^{i \Delta s\tau} - e^{i \Delta( s-1)\tau}\right] \right|.
    \label{eqn:generalDDerrorAPP}
\end{align}
According to the periodicity of the phase factors $e^{\pm i\Delta t}$,
the expression in Eq.~\eqref{eqn:generalDDerrorAPP} vanishes for any even number of segments $S \ge 4$.

\subsection{Second-order idle gate error \label{App:DDidle}}
We first evaluate the second-order contribution induced by XY crosstalk dynamics during idle operations, which serves as a reference for understanding the suppression achieved by dynamical decoupling. Under crosstalk dynamics, the second-order term in the Magnus expansion is
\begin{align}
     \bar{H}^{\mathrm{CD}(2)}_{\mathrm{idle}}=-\frac{i}{2T} \int _{0}^{T} \mathrm{d}t_{1} \int _{0}^{t_1} \mathrm{d}t_{2} \left[\tilde{H}_{\mathrm{XY}}(t_{1}),\tilde{H}_{\mathrm{XY}}(t_{2}) \right].
     \label{eqn:HCD2idle}
\end{align}

The commutator of the XY interaction at two different times $t_{1}$ and $t_{2}$ reads
\begin{align}
    &\left[\tilde{H}_{\mathrm{XY}}(t_{1}), \tilde{H}_{\mathrm{XY}}(t_{2}) \right]  \notag \\ &= \frac{J^{2}}{2} \left[ e^{i \Delta (t_{1}-t_{2}) } - e^{-i \Delta (t_{1}-t_{2}) } \right] \left( \hat{\sigma}_{1}^{z} \hat{I}_{2} -\hat{I}_{1} \hat{\sigma}_{2}^{z} \right).
   \label{eqn:Hxycommutator}
\end{align}
Substituting Eq.~\eqref{eqn:Hxycommutator} into Eq.~\eqref{eqn:HCD2idle}, we find that even at a matched gate time $T_{M} =2T_{\Delta}= 2\pi/\abs{\Delta}$, the second-order Magnus term under CD does not vanish and is given by
\begin{align}
    \bar{H}^{\mathrm{CD}(2)}_{\mathrm{idle}} =  \frac{J^2 }{2\Delta}\left( \hat{\sigma}_{1}^{z} \hat{I}_{2} -\hat{I}_{1} \hat{\sigma}_{2}^{z} \right).
    \label{eqn:H2CD}
\end{align}

We therefore identify Eq.~\eqref{eqn:H2CD} as the residual second-order error induced by XY crosstalk under CD. The corresponding second-order idle gate error is defined as the sum of the magnitudes of the coefficients of the second-order error terms,
\begin{align}
    \varepsilon^{\mathrm{CD}(2)}_{\mathrm{idle}}
    = 2 \abs{ \frac{J^2}{2\Delta}}.
    \label{eqn:eCD2idle}
\end{align}

We now turn to the evaluation of the second-order XY crosstalk error under dynamical decoupling in the idealized limit where the $Z_2$ pulses have zero width. During an idle period, the effective Hamiltonian defined in Eq.~\eqref{eqn:HDDeff} reduces to
\begin{align}
\tilde{H}^{\rm DD}_{\rm eff}(t)= f(t)\,\tilde{H}_{\mathrm{XY}}(t),
\end{align}
where $f(t)$ is a sign function $f(t)=(-1)^{s-1}$ for $t\in[(s-1)\tau,s\tau]$ that captures the sign inversion of the XY interaction within even DD segments.

The second-order Magnus term under DD is then given by
\begin{align}
&\bar{H}^{\mathrm{DD}(2)}_{\mathrm{idle}} \notag\\
&= -\frac{i}{2T}
\int_{0}^{T}\mathrm{d}t_{1}
\int_{0}^{t_{1}}\mathrm{d}t_{2}
\,f(t_{1})f(t_{2})
\big[\tilde{H}_{\mathrm{XY}}(t_{1}),
     \tilde{H}_{\mathrm{XY}}(t_{2})\big].
\label{eqn:HDD2idle_def}
\end{align}
For $t_{1}\in[(s_{1}-1)\tau,s_{1}\tau]$ and
$t_{2}\in[(s_{2}-1)\tau,s_{2}\tau]$, the product
$f(t_{1})f(t_{2})=(-1)^{s_{1}+s_{2}}$ encodes the relative sign of the XY
interaction between the two DD segments.

For the DD Z-4 sequence, the total gate time is divided into four equal segments.
Evaluating the time-ordered double integral in
Eq.~\eqref{eqn:HDD2idle_def} by summing over all segment combinations, we obtain,
at the matched gate time $T_{M}=2T_{\Delta}$,
\begin{align}
\bar{H}^{\mathrm{DD}(2)}_{\mathrm{idle}}
= \frac{\pi-4}{2\pi}\frac{J^{2}}{\Delta}
\left(
\hat{\sigma}_{1}^{z}\hat{I}_{2}
-\hat{I}_{1}\hat{\sigma}_{2}^{z}
\right).
\end{align}
The corresponding second-order idle-gate error is therefore
\begin{align}
\varepsilon^{\mathrm{DD}(2)}_{\mathrm{idle}}
= 2\left|\frac{\pi-4}{2\pi}\frac{J^{2}}{\Delta}\right|.
\label{eqn:eDD2idle}
\end{align}
Although the second-order term under DD does not vanish, its magnitude is reduced
compared with $\varepsilon^{\mathrm{CD}(2)}_{\mathrm{idle}}$, indicating that the DD
sequence suppresses the second-order contribution from XY crosstalk.

\subsection{Second-order $X_1$ gate error \label{App:DDX1}}
We now extend the second-order analysis to an active single-qubit operation by considering the $X_1$ gate under dynamical decoupling. As set up in Sec.~\ref{subsec:DDX1}, the $X_1$ gate is decomposed into two $\sqrt{X_1}$ rotations applied in the odd intervals between adjacent $Z_2$ gates within the DD Z-4 sequence. For analytical calculations of errors, we consider the $Z_2$ gate as an ideal gate with zero width. According to Eq.~\eqref{eqn:X_1DDgeneric} with $w=0$ and $\tau = T_M/4$, the drive Hamiltonian is given by 
\begin{align}
&\tilde{H}^{Q_1}_{\mathrm{drive}} (t)
    = \Omega_{1x}
      \cos\left(\frac{\pi}{\tau }
      \left[t - \left(s-\tfrac{1}{2}\right)\tau\right]\right)\hat{\sigma}_{1}^{x}, \notag \\
    &t \in [(s-1)\tau,\, s\tau],\quad s \in \{1,3\},
    \label{eqn:X_1DD}
\end{align}
and vanishes otherwise. The driving pulse amplitude is $\Omega_{1x} = \frac{\pi^{2}}{8\tau}$.

Under crosstalk dynamics, the second order term in the Magnus expansion of the $X_1$ gate evolution is given by
\begin{align}
    &\bar{H}^{\mathrm{CD}(2)}_{X_1} = -\frac{i}{2T}\int_{0}^{T}\ \mathrm{d}t_{1} \int_{0}^{t_1} \ \mathrm{d}t_{2} \left[ \tilde{H}^{\mathrm{CD}}(t_{1}),  \tilde{H}^{\mathrm{CD}}(t_{2}) \right],
     \label{eqn:HbarCD2X1}
\end{align}
where $\tilde{H}^{\mathrm{CD}}(t)=\tilde{H}_{\mathrm{XY}}(t)+\tilde{H}^{Q_1}_{\mathrm{drive}}(t)$. The commutator can be expanded as
\begin{align}
    &\left[\tilde{H}_{\mathrm{XY}}(t_{1}), \tilde{H}_{\mathrm{XY}}(t_{2}) \right] + \left[\tilde{H}^{Q_1}_{\mathrm{drive}}(t_1),\tilde{H}^{Q_1}_{\mathrm{drive}}(t_2) \right] + \notag \\ & \left[\tilde{H}_{\mathrm{XY}}(t_{1}), \tilde{H}^{Q_1}_{\mathrm{drive}}(t_2) \right] + \left[\tilde{H}^{Q_1}_{\mathrm{drive}}(t_1), \tilde{H}_{\mathrm{XY}}(t_{2}) \right].
    \label{eqn: HCDX1commutator}
\end{align}

The first component arises solely from the XY crosstalk and is identical to that in Eq.~\eqref{eqn:Hxycommutator}. The corresponding second-order contribution has already been evaluated in the idle-gate case. The second component corresponds to the ideal drive term and does not contribute to crosstalk-induced errors. Evaluating under the matched gate time $T_M = 2T_{\Delta} = 2\pi/|\Delta|$, we collect the resulting second-order error terms as
\begin{align}
    \frac{J^2 }{2\Delta}\left( \hat{\sigma}_{1}^{z} \hat{I}_{2} -\hat{I}_{1} \hat{\sigma}_{2}^{z} \right) - \frac{J}{4}\left( \hat{\sigma}_{1}^{z} \hat{\sigma}_{2}^{-}  +\hat{\sigma}_{1}^{z} \hat{\sigma}_{2}^{+} \right).
\end{align}
The first part arises from the XY crosstalk, as shown in Eq.~\eqref{eqn:H2CD}, while the second part originates from the commutator between the XY crosstalk and the drive.
Accordingly, the second-order error $ \varepsilon^{\mathrm{CD}(2)}_{X_1}$ of the CD is quantified as the sum of the magnitudes of the coefficients of all second-order error terms.
\begin{align}
    \varepsilon^{\mathrm{CD}(2)}_{X_1}= 2\abs{\frac{J^2 }{2\Delta}} + 2 \abs{\frac{J}{4}}.
    \label{eqn:eCD2X1}
\end{align}

We now turn to the evaluation of the second-order contribution under dynamical decoupling. During the $X_1$ gate under DD Z-4 sequence, the effective Hamiltonian takes the form as in Eq.~\eqref{eqn:HDDeff},
\begin{align}
\tilde{H}^{\rm DD}_{\rm eff}(t)=
f(t)\tilde{H}_{\mathrm{XY}}(t) + \tilde{H}^{Q_1}_{\rm drive}(t),
\end{align}
where $\tilde{H}^{Q_1}_{\mathrm{drive}}(t)$ is defined in Eq.~\eqref{eqn:X_1DD} and vanishes outside the odd segments, with $f(t)=(-1)^{s-1}$ for $t\in[(s-1)\tau,s\tau]$.

Similar to the crosstalk dynamics case, we evaluate the second order term in the Magnus expansion at the matched gate time $T_M=2T_{\Delta}=2\pi/|\Delta|$ and collect the resulting error terms as
\begin{align}
    \bar{H}^{\mathrm{DD}(2)}_{X_1} =&   \frac{\pi-4}{2\pi}\frac{J^2 }{\Delta}\left( \hat{\sigma}_{1}^{z} \hat{I}_{2} -\hat{I}_{1} \hat{\sigma}_{2}^{z} \right) \notag \\&+ \frac{iJ}{4}\left( \hat{\sigma}_{1}^{z} \hat{\sigma}_{2}^{-}  -\hat{\sigma}_{1}^{z} \hat{\sigma}_{2}^{+} \right).
\end{align}
The first term originates from the XY crosstalk under dynamical decoupling, while the second term arises from the commutator between the XY interaction and the drive.

Accordingly, the second-order $X_1$-gate error under DD is defined as the sum of the magnitudes of the coefficients of all second-order error terms,
\begin{align}
\varepsilon^{\mathrm{DD}(2)}_{X_1}
= 2\abs{\frac{\pi-4}{2\pi}\frac{J^{2}}{\Delta}}
+ 2\abs{\frac{J}{4}}.
\label{eqn:eDD2X1}
\end{align}
Although the second-order contribution does not vanish, dynamical decoupling reduces its magnitude, indicating suppression of XY-crosstalk-induced errors during the $X_1$ gate operation.

\subsection{Single-site scheme of dynamical decoupling \label{app:SingleQDD}}

\begin{figure}[htbp]
\begin{center}
\includegraphics[width=1\linewidth]{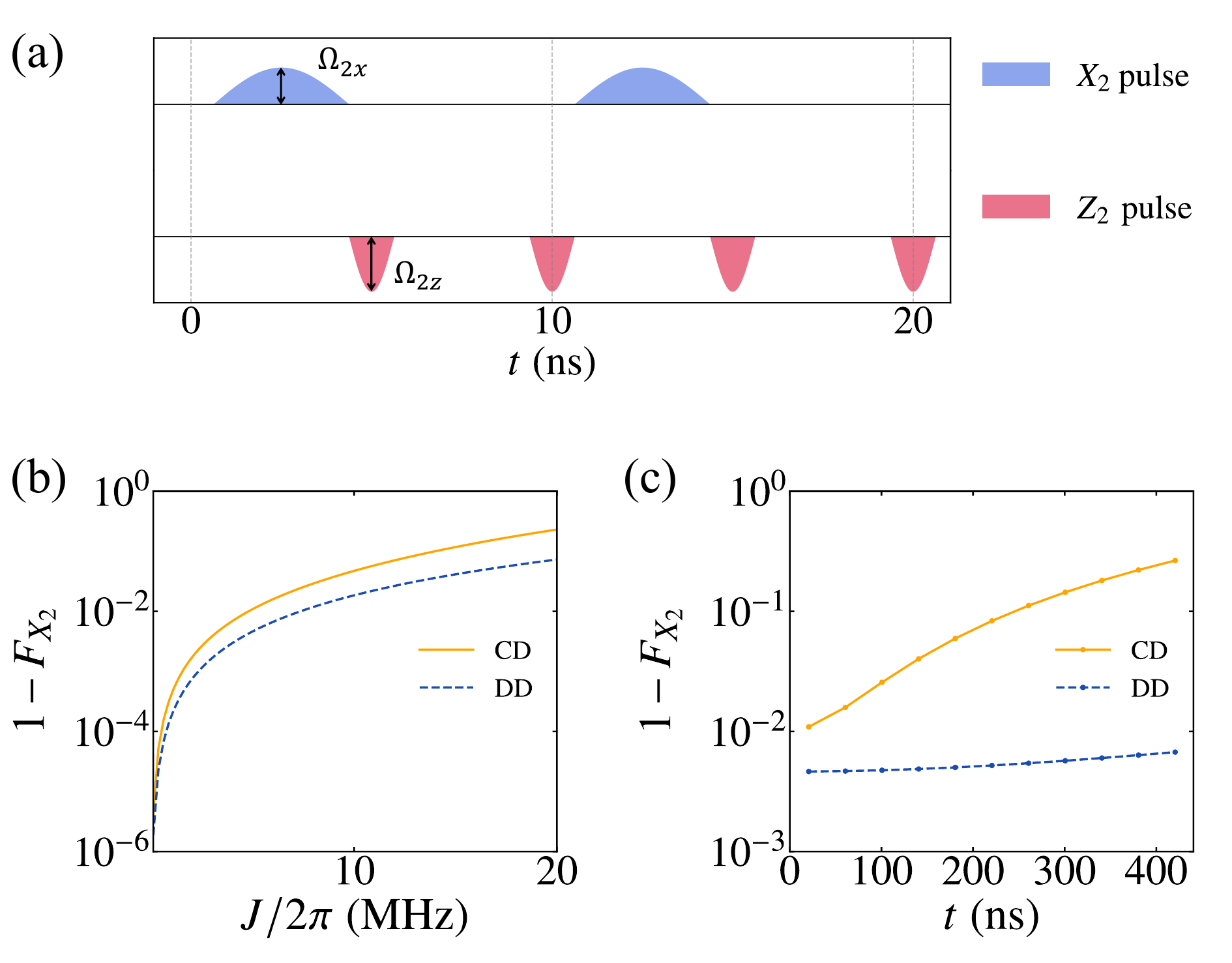}
\caption{Suppression of $X_2$ gate infidelity under single-site dynamical decoupling.
(a) Waveforms of the $X_2$ control pulses (generated by two $\sqrt{X_2}$ pulses) and the DD Z-4 sequence containing four $Z_2$ gates applied to $Q_2$.
(b) $X_{2}$ gate infidelity of one $X_2$ operation as a function of the coupling strength $J$ under DD Z-4.
(c) $X_{2}$ gate infidelity of consecutive $X_2$ operations as a function of time under DD Z-4, with $J/2\pi = 5$~MHz.}
    \label{fig:DDX2Infidelity}
\end{center}
\end{figure}

We next consider the case where both the control drive and the DD Z-4 sequence are applied to the same qubit, $Q_{2}$, enabling the simultaneous application of single-qubit operations and suppression of crosstalk on a single site. The single-qubit $X_{2}$ gate is realized through a driving Hamiltonian of the form
\begin{align}
&\tilde{H}^{Q_2}_{\mathrm{drive}} (t)
    = \Omega_{2x}(w)
      \cos\left(\frac{\pi}{\tau - w}
      \left[t - \left(s-\tfrac{1}{2}\right)\tau\right]\right)\hat{\sigma}_{2}^{x}, \notag \\
    &t \in [(s-1)\tau+\tfrac{w}{2},\, s\tau-\tfrac{w}{2}],\quad s \in \mathrm{odd},
    \label{eqn:X_2DDgeneric}
\end{align}
where the driving pulse amplitude is $\Omega_{2x}(w) = \frac{\pi^{2}}{8(\tau - w)}$, and $w$ denotes the pulse width of the $Z_2$ gate. The driving pulse shape is identical to that in Eq.~\eqref{eqn:X_1DDgeneric}. As a result, the second-order crosstalk error $\varepsilon^{\mathrm{DD}(2)}_{X_2}$ is identical to $\varepsilon^{\mathrm{DD}(2)}_{X_1}$, since the same DD-based driving protocol is applied to $Q_{2}$. An example of the corresponding waveforms with $w=\tau/4$ is shown in Fig.~\ref{fig:DDX2Infidelity}(a).

For $\Delta/2\pi = 50$~MHz and a matched gate time of $T_{M} = 20$~ns, as shown in Fig.~\ref{fig:DDX2Infidelity}(b), the DD sequence reduces the $X_{2}$-gate infidelity by approximately $0.37$ orders of magnitude compared with the crosstalk-dynamics case. Under repeated gate operations, the suppression effect remains stable: after 21 consecutive $X_{2}$ gates, the infidelity remains more than one order of magnitude lower than that without DD, as shown in Fig.~\ref{fig:DDX2Infidelity}(c). These results are consistent with those presented in Fig.~\ref{fig:DDX1Infidelity}, confirming that the control drive and the DD Z-4 sequence can be applied to the same qubit. This establishes the feasibility of integrating active gate control and dynamical decoupling on a single qubit to suppress crosstalk-induced errors.

\subsection{Parallel $X_1X_2$ operations under dynamical decoupling \label{app:X1X2DD}}
\begin{figure}[htbp]
    \begin{center}
    \includegraphics[width=1\linewidth]{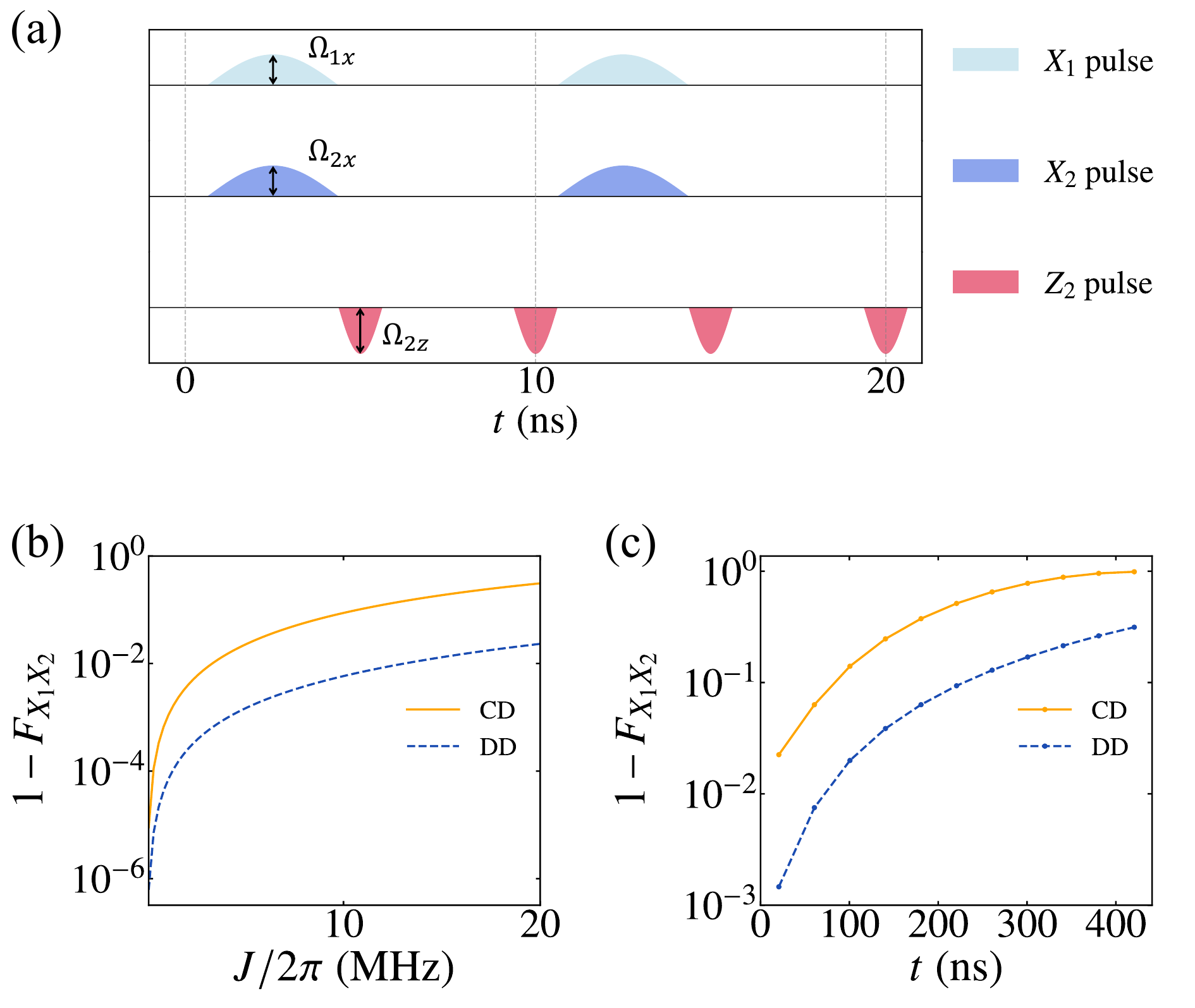} \caption{Suppression of parallel $X_{1}X_{2}$ gate infidelity under dynamical decoupling.(a) Waveforms of the parallel $X_1$ and $X_2$ control pulses (each generated by two $\sqrt{X}$ pulses) and the DD Z-4 sequence containing four $Z_2$ gates applied to $Q_2$. (b) Parallel $X_{1}X_{2}$ gate infidelity of one operation as a function of the coupling strength $J$ under DD Z-4. (c) Parallel $X_{1}X_{2}$ gate infidelity of consecutive operations as a function of time under DD Z-4, with $J/2\pi=5$~MHz.}
    \label{fig:DDX1X2infidelity}
    \end{center}
\end{figure}

Building upon the previous results on XY-crosstalk suppression during $X_{2}$ gates under the DD Z-4 sequence, we extend our analysis to the case where parallel $X_{1}X_{2}$ gates are applied simultaneously to both qubits. In this configuration, the $X_{1}$ gate is applied to $Q_1$, while the DD Z-4 sequence together with the $X_{2}$ gate is applied to $Q_2$, as illustrated in Fig.~\ref{fig:DDX1X2infidelity}(a).

To quantify the crosstalk-induced error during this parallel operation, we evaluate the second-order crosstalk error under DD as
\begin{align}
     \varepsilon^{\mathrm{DD}(2)}_{X_1 X_2} &= \varepsilon^{\mathrm{DD}(2)}_{X_1} + \varepsilon^{\mathrm{DD}(2)}_{X_2} -\varepsilon^{\mathrm{DD}(2)}_{\mathrm{idle}} \notag \\ &= 2\abs{ \frac{\pi-4}{2\pi}\frac{J^2 }{\Delta}} + 4\abs{\frac{J}{4}}.
\end{align}
In the absence of DD, the corresponding second-order error under crosstalk dynamics, $\varepsilon^{\mathrm{CD}(2)}_{X_1 X_2}$, is given by
\begin{align}
    \varepsilon^{\mathrm{CD}(2)}_{X_1 X_2} &= \varepsilon^{\mathrm{CD}(2)}_{X_1} + \varepsilon^{\mathrm{CD}(2)}_{X_2} -\varepsilon^{\mathrm{CD}(2)}_{\mathrm{idle}} \notag \\ &= 2\abs{\frac{J^2}{2\Delta}} + 4\abs{\frac{J}{4}}.
\end{align}
A direct comparison shows that $\varepsilon^{\mathrm{DD}(2)}_{X_1 X_2} < \varepsilon^{\mathrm{CD}(2)}_{X_1 X_2}$, confirming that the DD sequence effectively suppresses XY crosstalk even during simultaneous parallel operations.

For the parallel $X_{1}X_{2}$ gate configuration, the dependence of gate infidelity on the coupling strength $J$ is shown in Fig.~\ref{fig:DDX1X2infidelity}(b). When $J/2\pi \geq 5$~MHz, the application of DD reduces the gate infidelity by approximately $1.2$ orders of magnitude. This suppression is stronger than that observed for a single $X_1$ gate applied to $Q_1$ alone [see Fig.~\ref{fig:DDX1Infidelity}(b)].

We further examine the performance of DD under repeated parallel gate operations. As shown in Fig.~\ref{fig:DDX1X2infidelity}(c), odd numbers of parallel $X_{1}X_{2}$ gate sets are applied. At early times, the infidelity is reduced by more than one order of magnitude. However, in contrast to the single-$X$-gate case [Fig.~\ref{fig:DDX1Infidelity}(c)], DD does not effectively suppress the subsequent growth of infidelity under repeated parallel $X_{1}X_{2}$ operations. After up to 21 gate sets, the net reduction in infidelity is limited to approximately half an order of magnitude.

\bibliography{2025XYCrossTalk}
\end{document}